\documentclass[journal]{IEEEtran}
\IEEEoverridecommandlockouts
\usepackage{bbding}
\usepackage{url}
\usepackage{booktabs}

\usepackage{CJK}
\usepackage{subfigure}
\usepackage{amsmath,amssymb}
\usepackage{multirow}
\usepackage{fancybox}
\usepackage[dvipsnames,svgnames,x11names]{xcolor}
\usepackage{graphicx}
\usepackage{bbm}
\usepackage{algorithm} 
\usepackage{algorithmic} 
\usepackage{verbatim}
\usepackage{booktabs}
\usepackage[UKenglish]{babel}
\usepackage{setspace}
\usepackage{color}
\usepackage{lipsum}
\usepackage{mathtools, cuted}
\usepackage{nomencl}
\usepackage{multicol}

\makenomenclature

\makeatletter
\newcommand*{\rom}[1]{\expandafter\@slowromancap\romannumeral #1@}
\makeatother

\DeclareMathOperator*{\argmax}{argmax}

\ifCLASSINFOpdf
\else
\fi
\begin{document}

%
\title{Stochastic Dynamic Pricing for EV Charging Stations with Renewable Energy Integration and Energy Storage}
%
%
%

\author{Chao~Luo,
        Yih-Fang~Huang,~\IEEEmembership{Fellow,~IEEE,}
        and~Vijay~Gupta,~\IEEEmembership{Member,~IEEE}
\thanks{C. Luo, Y.-F. Huang, and V. Gupta are with the Department
of Electrical Engineering, University of Notre Dame, Notre Dame,
IN, 46556 USA e-mail: \{cluo1, huang, vgupta2\}@nd.edu }.
\thanks{This paper was presented, in part, at the 2nd International Conference on Vehicle Technology and Intelligent Transport Systems. This work has been partially supported by the National Science Foundation under grants CNS-1239224, ECCS-1550016, and CPS-1544724.}
}

\maketitle

\begin{abstract}
This paper studies the problem of stochastic dynamic pricing and energy management policy for electric vehicle (EV) charging service providers. In the presence of renewable energy integration and energy storage system, EV charging service providers must deal with multiple uncertainties --- charging demand volatility, inherent intermittency of renewable energy generation, and wholesale electricity price fluctuation. The motivation behind our work is to offer guidelines for charging service providers to determine proper charging prices and manage electricity to balance the competing objectives of improving profitability, enhancing customer satisfaction, and reducing impact on power grid in spite of these uncertainties. We propose a new metric to assess the impact on power grid without solving complete power flow equations. To protect service providers from severe financial losses, a safeguard of profit is incorporated in the model. Two algorithms --- stochastic dynamic programming (SDP) algorithm and greedy algorithm (benchmark algorithm) --- are applied to derive the pricing and electricity procurement policy. A Pareto front of the multi-objective optimization is derived. Simulation results show that using SDP algorithm can achieve up to 7\% profit gain over using greedy algorithm. Additionally, we observe that the charging service provider is able to reshape spatial-temporal charging demands to reduce the impact on power grid via pricing signals.
\end{abstract}



%

\section*{Nomenclature}
\addcontentsline{toc}{section}{Nomenclature}
\begin{IEEEdescription}[\IEEEusemathlabelsep\IEEEsetlabelwidth{$J(I_k,u_k)$}]
\item[$K$] Total number of planning horizons.
\item
\item[$N$] Number of buses in a power network.
\item
\item[$M$] Number of PQ buses in a power network.
\item
\item[$s_j$] The $j$-th charging station.
\item
\item[$p_{kj}$] Charging price of the $j$-th charging station at the $k$-th horizon.
\item
\item[$d_{kj}$] Charging demand of the $j$-th charging station at the $k$-th horizon.
\item
\item[$c_k$] Real time wholesale electricity price at the $k$-th horizon.
\item
\item[$E$] Electricity storage capacity.
\item
\item[$I_k$] Remaining electricity in storage at the beginning of the $k$-th horizon.
\item
\item[$W_k$] Profit at the $k$-th horizon.
\item
\item[$W_{\textrm{min}}$] Threshold for profit safeguard.
\item
\item[$G_k$] Customer satisfaction at the $k$-th horizon.
\item
\item[$F_k$] Impact on power grid at the $k$-th horizon.
\item
\item[$o_k$] Electricity purchase at the $k$-th horizon.
\item
\item[$u_k$] Renewable energy at the $k$-th horizon.
\item
\item[$\eta_s$] Unit storage cost (measured in \$/MWh).
\item
\item[$\eta_c$] Charging efficiency.
\item
\item[$\eta_d$] Discharging efficiency.
\item
\item[$\alpha$] Shape parameter in customer satisfaction formula.
\item
\item[$\omega$] Shape parameter in customer satisfaction formula.
\item
\item[$\phi_k$] Total charging demand at the $k$-th horizon.
\item
\item[$\Pi_k$] Total utility at the $k$-th horizon.
\item
\item[$\lambda_1$] Profit weight coefficient.
\item
\item[$\lambda_2$] Customer satisfaction weight coefficient.
\item
\item[$\lambda_3$] Impact weight coefficient.
\item
\item[$\gamma_{i,j}$] Price elasticity coefficient.
\item
\item[$P_i$] Active power of the $i$-th bus.
\item
\item[$Q_i$] Reactive power of the $i$-th bus.
\item
\item[$v_i$] Voltage magnitude of the $i$-th bus.
\item
\item[$\delta_i$] Voltage phase of the $i$-th bus.
\item
\item[$G_{ik}$] Conductance of the $ik$-th element of the bus admittance matrix.
\item
\item[$B_{ik}$] Susceptance of the $ik$-th element of the bus admittance matrix.
\item
\item[$S_i^{\textrm{Ac}}$] Active power sensitivity of the $i$-th bus.
\item
\item[$S_i^{\textrm{Re}}$] Reactive power sensitivity of the $i$-th bus.
\item
\item[$J(I_k,u_k)$] Maximum expected aggregated utility from the $k$-th horizon to the $k$-th horizon.
\end{IEEEdescription}

\section{Introduction}
Electric vehicles (EVs) exhibit many advantages over fossil fuel driven vehicles in terms of operation and maintenance cost, energy efficiency, and gas emission \cite{asimpson}-\cite{rsioshansi}. However,  the fear of limited driving distance (range anxiety) is hanging over EV drivers' heads like the Sword of Damocles. To alleviate this range anxiety, the capacity of on-board battery should be increased and more EV charging stations should be deployed. Intensive research work has been carried out to study how to strategically deploy charging stations \cite{cluo1}-\cite{yli}. Currently, EV charging service is primarily provided for free as one of the employee benefits in some organizations or as a perk to those owners of some specific EV models (e.g. Tesla). There is a lack of viable and profitable pricing and energy management model for public charging stations. Our goal is to offer guidelines for charging service providers to make informed and insightful decisions on pricing and electricity procurement by jointly optimizing multiple objectives under uncertainties.

There is a growing literature aimed at providing guidelines for economic operation of EV charging stations. In \cite{yguo1}-\cite{yguo2}, the authors studied a dynamic pricing scheme to improve the revenue of an EV parking deck. However, their model did not take into account customer satisfaction and the impact on power grid due to EV charging. In \cite{jfoster}-\cite{sbeer}, several algorithms have been proposed for a power aggregator to manage EV charging loads and submit bids to electricity market to provide regulation service (RS). Game theory based approaches have been used to model the interplay among multiple EVs or between EVs and power grid in \cite{yhan}-\cite{rcouillet}.  Yan \emph{et al.} presented a multi-tier real time pricing algorithm for EV charging stations to encourage customers to shift their charging schedule from peak period to off-peak period \cite{qyan}. Nevertheless, they did not consider that some customers may strategically change their charging schedule in response to pricing signals. In \cite{dban}, Ban \emph{et al.} employed multi queues to model the arrivals and departures of EVs among multiple charging stations. Pricing signals were used to guide EVs to different charging stations to satisfy the predefined quality of service (QoS); but the interactions between EV charging and power grid was not analysed in their model.  A distributed network cooperative method was proposed to minimize the charging cost of EVs while guaranteeing that the aggregated load satisfies safety limits \cite{nrahbariasr}. Their model, however, did not incorporate renewable energy generation and consider charging demand volatility.

In our model, we take a comprehensive view of these interweaving issues pertaining to EV charging pricing and energy management. Specifically, we formulate our problem to simultaneously optimize multiple objectives --- improving the profit, enhancing the customer satisfaction, and reducing the impact on power grid in the light of renewable energy generation and energy storage. Our model takes into account multiple uncertainties including charging demand volatility, inherent intermittency of renewable energy generation, and real time wholesale electricity price fluctuation. For each type of uncertainty, an appropriate model is proposed and incorporated in the overall optimization framework. Finally, a stochastic dynamic programming (SDP) algorithm is employed to derive the charging prices and the electricity procurement from the power grid for each planning horizon. Besides, SDP algorithm has been used for water reservoir operation in \cite{cozelkan}-\cite{akerr}.  In terms of the electricity retail market, a game theory based dynamic pricing scheme is studied in \cite{ljia} which also takes into account renewable integration and local storage.

The main contributions of our work are as follows:
\begin{itemize}
\item We proposed a multi-objective optimization framework to solve the problem, and the solutions provide us insights into how to make a tradeoff among multiple objectives of the profitability, the customer satisfaction, and impact on power grid, and offer guidance to set charging prices to balance the charging demand across the power system.
\item We used Newton's method to derive a fast-computing metric to assess the impact of EV charging on power grid, which frees us from solving the complete nonlinear power flow equations. This metric also can be used to analyze other electric load's impact on power grid.
\item We derived the active power and reactive power sensitivities for the load buses in a power system which can serve as a guideline for EV charging station placement to alleviate the charging stress on the power grid.
\item In terms of market risk, we introduced a safeguard of profit for EV charging service providers, which raises a warning when the profit is likely to reach a dangerous threshold. This mechanism is beneficial for the charging service provider to safely manage its capital and avoid severe financial losses.
\end{itemize}

The remainder of the paper is organized as follows: Section \rom{2} presents the general problem formulation. Section \rom{3} introduces charging demand estimation and Section \rom{4} discusses how to assess the impact on power grid from EV charging. Renewable energy and real time wholesale electricity price forecast, the safeguard of profit, and SDP algorithm are introduced in Section \rom{5}. Section \rom{6} presents simulation results and discussions. Conclusions are provided in Section \rom{7}. A nomenclature table is also provided as a reference.

\section{Problem Formulation}
\begin{figure}[htbp]
\centering
\includegraphics[width=3in]{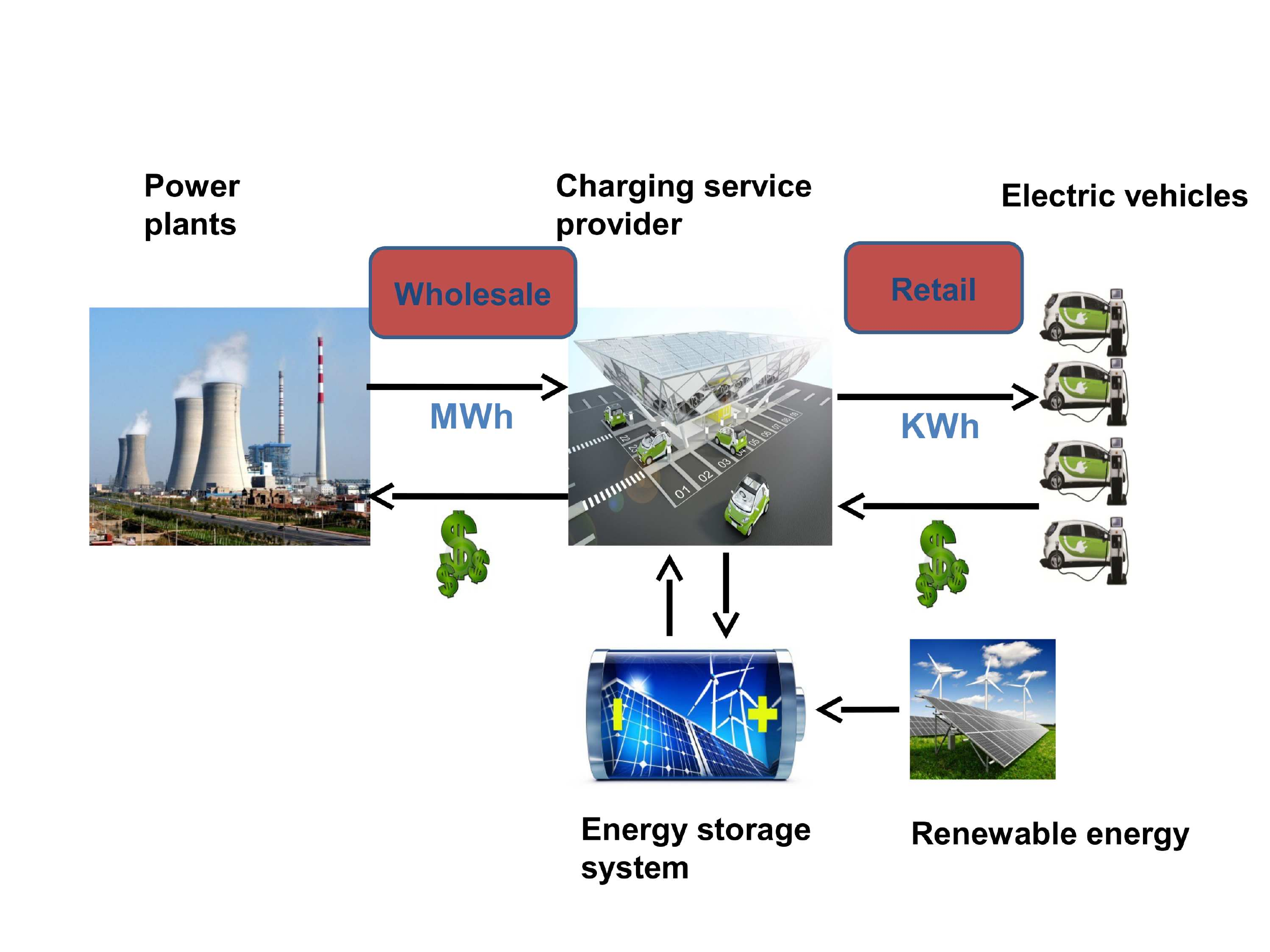}
\caption{Business Model of EV Charging Service Provider}
\label{fig1}
\end{figure}

In this study, we assume that an EV charging service provider operates a set of charging stations within a large region. As a mediator between the wholesale market and end customers (EVs), the charging service provider procures electricity from the wholesale market and resells it to EVs. We also assume that the service provider is able to harvest renewable energy (i.e. solar or wind power) and save it in an energy storage system. An overview of the EV charging service provider's model is illustrated in Fig. \ref{fig1}.

\subsection{Profit of Charging Service Provider}
In the United States, the Independent System Operator (ISO) or the Regional Transmission Organization (RTO) collects supply offers from power plants and demand bids from load serving entities (LSEs) or market participants, calculates the day-ahead wholesale prices and real time spot prices, coordinates and monitors the economic dispatch of electricity across a vast region \cite{rhuisman}-\cite{mventosa}. We assume that the charging service provider is an LSE, who purchases electricity from the wholesale real time market and resells it to EVs. Let $\mathcal{S}=\{s_1,s_2,\cdots,s_L\}$ denote the charging stations operated by the service provider. A day is divided into $K$ planning horizons. At the start of each horizon, the service provider will publish new charging prices during this horizon. Price differentiation is allowed across charging stations.  Let $\mathcal{P}=\{p_{k1},p_{k2},\cdots,p_{kL}\},k=1,2,\cdots,K$ denote the charging prices in the $k$-th horizon, and $o_k$ denote electricity procurement from the wholesale real time market. We use wholesale real time electricity market prices in our theoretical analysis. Let $\mathcal{C}=\{c_1,c_2,\cdots,c_K\}$ represent wholesale real time electricity prices. In addition, we assume that the service provider has an energy storage system with capacity $E$ MWh. Let $I_k$ denote the electricity in the storage at the beginning of the $k$-th horizon, and $u_k$ be the renewable energy generation during the $k$-th horizon. The profit made in the $k$-th horizon is given by

\begin{equation}
\begin{aligned}
W_k&=\sum_{j=1}^Lp_{kj}d_{kj}-c_ko_k-\\
&\eta_s(I_k+\eta_cu_k+\eta_co_k-\frac{1}{\eta_d}\sum_{j=1}^Ld_{kj} + w_k),
\end{aligned}
\end{equation}
where $d_{kj}$ corresponds to the charging demand (electricity consumption) at the $j$-th station in the $k$-th horizon, $\sum_{j=1}^Lp_{kj}d_{kj}$ is the total revenue, $c_ko_k$ is the cost of electricity procurement, and $\eta_s$(\$/MWh) is the unit storage cost, which includes capital cost and maintenance cost. Besides, $\eta_c$ ($0<\eta_c<1$) and $\eta_d$ ($0<\eta_d<1$) are charging efficiency and discharging efficiency, respectively. And $w_k$ is the process noise of the energy storage system, which has a Gaussian distribution with zero mean and variance $\sigma_w^2$.

\subsection{Customer Satisfaction}
\begin{figure}[htbp]
\centering
\includegraphics[width=3in]{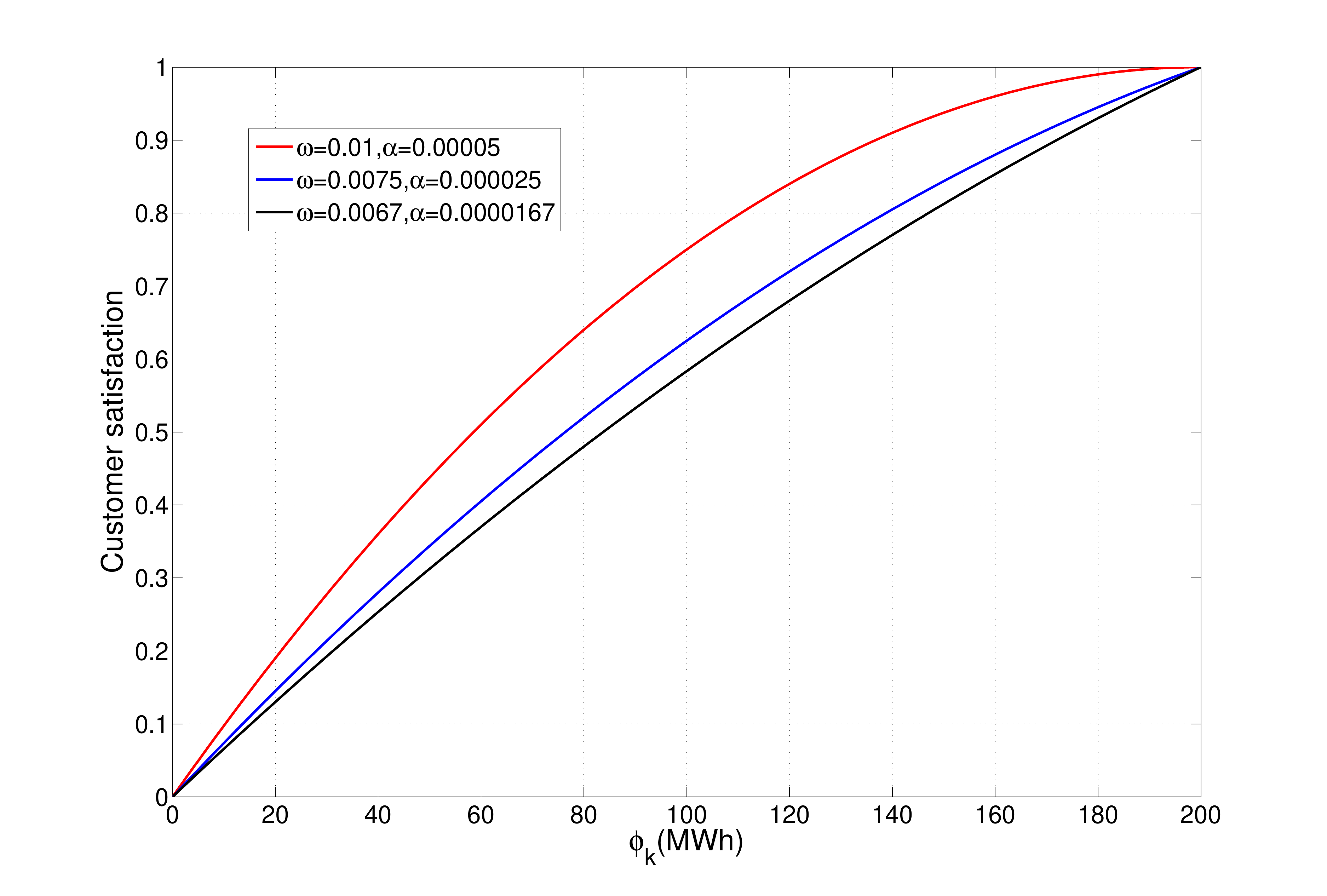}
\caption{Sample Customer Satisfaction Functions ($E=200$)}
\label{fig2}
\end{figure}

Customer satisfaction helps to build up customer loyalty, which can reduce the efforts to allocate market budgets to acquire new customers. Poor customer satisfaction will discourage people to purchase EVs, affecting the development of entire EV industry. Customer satisfaction is one of the objectives in our multi-objective optimization framework. Several customer satisfaction evaluation methods have been investigated in \cite{pyang}-\cite{rfaranda}. In this paper, we consider the market-level customer satisfaction instead of the individual-level satisfaction. We use a quadratic function to formulate the overall customer satisfaction of all EVs in a horizon, namely,

\begin{equation}
\label{satisfactioneq}
G_k=-\frac{\alpha}{2}\phi_k^2+\omega\phi_k,\;0\leq \phi_k \leq E
\end{equation}
where $E$ is the electricity storage capacity, $\omega$ and $\alpha$ are shape parameters, $\phi_k$ is the aggregated charging demand (electricity consumption) of all EVs in the $k$-th horizon which is defined as,

\begin{equation}
\phi_k=\sum_{j=1}^Ld_{kj}.
\end{equation}
Eq. (\ref{satisfactioneq}) with different shape parameters is plotted in Fig. \ref{fig2}. In plotting Fig. \ref{fig2}, we choose the shape parameters $\alpha$ and $\omega$ such that the concave function $G_k$ has a minimum of 0, which indicates that EV drivers have the least satisfaction, and a maximum of 1, which indicates that they have the most satisfaction. Note that Eq. (\ref{satisfactioneq}) is a non-decreasing function with a non-increasing first order derivative. This implies that customer satisfaction will always grow as the total charging demand $\phi_k$ increases, but the growth rate will decrease and customer satisfaction tends to get saturated as the total charging demand approaches the storage capacity $E$. This is a standard assumption following the law of diminishing marginal utility (Gossen's First Law) in economics \cite{gheinrich}.

\subsection{Impact on Power Grid}
Large-scale EV charging presents a substantial load to power networks \cite{hturker}-\cite{kkumar}. Many studies have shown that uncoordinated EV charging can affect the normal operation of power grid in terms of severe power loss, voltage variation, frequency deviation, and harmonics problems \cite{jlopes}-\cite{renac}. Usually, grid frequency can be well maintained either by the power generator side using automatic gain control (AGC) \cite{pkundur} or by the load side using certain demand response techniques \cite{czhao}-\cite{mmoghadam}. In our study here, we only consider the impact of voltage variation (magnitude and phase). In addition, we assume that a higher-level entity like an aggregator or ISO/RTO can take care of network transmission constraint issues within the power system under its supervision, so the EV charging service provider does not need to worry about transmission constraint problem. Let $F_k$ denote the impact of EV charging on power grid at the $k$-th horizon.

\begin{equation}\label{impact}
F_k=f(d_{k1},d_{k2},\cdots,d_{kL}),
\end{equation}
where $d_{kj}$ is the charging demand at the $j$-th charging station in the $k$-th horizon, and $f(.)$ is a function to be discussed in Section \rom{4}. Function $f(.)$ should reflect the basic assumption that the impact on power grid increases when the charging demands increase.

\subsection{Multi-objective Optimization Framework}
A multi-objective optimization problem arises naturally from the fact that the charging service provider needs to balance multiple competing objectives --- maximizing profit, maximizing customer satisfaction, and minimizing the impact on power grid. For the $k$-th horizon, we formulate the multi-objective optimization as follows,

\begin{equation}\label{wholeproblem}
\begin{aligned}
&\max_{\mathbf{X}_k}\left\{\mathbb{E}(W_k),\mathbb{E}(G_k),\mathbb{E}(-F_k)\right\}\\
&\;\;\;\;\textrm{s.t.  } \mathbf{X}_k\in U(\mathbf{X}_k),
\end{aligned}
\end{equation}
where $\mathbf{X}_k=[p_{k1},p_{k2},\cdots,p_{kL},o_k]^{\textrm{T}}$ is the vector of decision variables, and $\mathbb{E}(.)$ represents the expectation operation.

Clearly, there are several approaches to solve multi-objective optimization problems: weighted sum approach, adaptive weighted sum approach, $\epsilon$-constraint approach, a priori approach and a posteriori approach \cite{clhwang}-\cite{ikim}, among others. The weighted sum approach is not suitable for obtaining the whole Pareto front if the main objective function is not convex. In this paper, we use an adaptive weighted sum approach discussed in \cite{ikim} to derive the Pareto front of Eq. (\ref{wholeproblem}). The main idea of the adaptive weighted sum approach is that firstly we use the ordinary weighted sum to obtain the basic shape of Pareto front, and then refine it by recursively reducing mesh size within the Pareto front. First, we rewrite the problem as follows,

\begin{equation}
\scalebox{0.91}{$
\begin{aligned}
&\max_{\mathbf{X}_k}\mathbb{E}(\Pi_k)=\mathbb{E}\left\{\lambda_1\frac{W_k}{W_k^{\textrm{max}}}+\lambda_2\frac{G_k}{G_k^{\textrm{max}}}
-\lambda_3\frac{F_k}{F_k^{\textrm{max}}}\right\}\\
&\;\;\;\;\textrm{s.t.  } \mathbf{X}_k\in U(\mathbf{X}_k),
\end{aligned}$}
\end{equation}
where $\lambda_1,\lambda_2,$ and $\lambda_3$ are nonnegative coefficients, satisfying the constraint of $\lambda_1+\lambda_2+\lambda_3=1$.  Different weight vectors ($\lambda_1, \lambda_2, \lambda_3$) generates different convex Pareto optima. The non-convex part of this Pareto front can be found in the refinement phase. Additionally, $W_k^{\textrm{max}}$, $G_k^{\textrm{max}}$, and $F_k^{\textrm{max}}$ are the maximum values of each objective function in the $k$-th horizon.

Our ultimate goal is to maximize the aggregated utility across multiple horizons.

\begin{equation}\label{eq1}
\begin{aligned}
&(\mathbf{X}_1^*,\cdots,\mathbf{X}_K^*)=\argmax_{\mathbf{X}_1,\cdots,\mathbf{X}_K}
\left\{\sum_{k=1}^K\mathbb{E}(\Pi_k)\right\},\\
&\;\;\;\;\;\;\textrm{s.t.  } \mathbf{X}_k\in U(\mathbf{X}_k), k=1,2,\cdots,K.
\end{aligned}
\end{equation}

To solve this multi-horizon and multi-objective optimization problem, we face several challenges: (1) How do we accurately estimate the charging demand $d_{kj}$ at each charging station? (2) How do we develop an appropriate metric to assess the impact on power grid defined in Eq. (\ref{impact})? (3) How should we incorporate a safeguard of profit to prevent severe financial losses? (4) How can we solve this complex optimization problem in an efficient manner? In the following sections, we will address these challenges in details.

\section{Charging Demand Estimation}
In practice, EV drivers will adjust their charging demands and charging schedules in response to charging prices. The charging demand function $d_{kj}$ thus should characterize customers' response to price fluctuations. In our work, an online linear regression model \cite{rchristensen}-\cite{adobson} is employed to predict the charging demand $d_{kj}$. For each charging station, the predicted charging demand is defined as

\begin{equation}\label{lrm}
\scalebox{0.9}{$
\begin{cases}
d_{k1}=\gamma_{0,1}-\gamma_{1,1}p_{k1}+\gamma_{2,1}p_{k2}+\cdots+\gamma_{L,1}p_{kL}+\epsilon_{k1},\\
d_{k2}=\gamma_{0,2}+\gamma_{1,2}p_{k1}-\gamma_{2,2}p_{k2}+\cdots+\gamma_{L,2}p_{kL}+\epsilon_{k2},\\
\vdots\\
d_{kL}=\gamma_{0,L}+\gamma_{1,L}p_{k1}+\gamma_{2,2}p_{k2}+\cdots-\gamma_{L,L}p_{kL}+\epsilon_{kL},\\
\end{cases}$}
\end{equation}
where $\gamma_{0,j}(j=1,2,\cdots, L)$ is the intercept of the $j$-th linear regression equation, $\gamma_{i,j}=\gamma_{j,i}(i\neq j)$ are the cross-price elasticity coefficients, reflecting how the change of the charging price at station $j$ can influence the charging demand at station $i$, and $\gamma_{i,i}$ is the self-price elasticity coefficient, reflecting how the change of the charging price of station $i$ can influence its own charging demand. Finally $\epsilon_{kj}(j=1,2,\cdots,L)$ is assumed to be an independent Gaussian random variable with mean 0 and variance $\sigma_{kj}^2$. The variable $\epsilon_{kj}$ captures the unknown random charging demand which cannot be characterized by the linear terms.

Recursive least square (RLS) algorithm is a common method applied to estimate the coefficients in Eq. (\ref{lrm}) using historical data \cite{jproakis}-\cite{dbertsimas}. Let $\mathbf{Y}_j=[\gamma_{0,j},\gamma_{1,j},\cdots,\gamma_{L,j}]^{\textrm{T}}$ denote the vector of price elasticity coefficients related to the $j$-th charging station. Applying RLS, we have the following update equations,

\begin{equation}\label{update1}
\begin{aligned}
\begin{cases}
&e_{kj}=d_{kj}-\mathbf{P}_k^{\textrm{T}}\mathbf{Y}_j,\\
&g_{kj}=\frac{\mathbf{H}_{(k-1)j}\mathbf{P}_k}{\nu+\mathbf{P}_{k}^{\textrm{T}}\mathbf{H}_{(k-1)j}\mathbf{P}_k},\\
&\mathbf{H}_{kj}=\nu^{-1}\mathbf{H}_{(k-1)j}-g_{kj}\mathbf{P}_k^{\textrm{T}}\nu^{-1}\mathbf{P}_k,\\
&\mathbf{Y}_j\leftarrow \mathbf{Y}_j+e_{kj}g_{kj},\\
\end{cases}
\end{aligned}
\end{equation}
where $\nu$ is the forgetting factor. Besides, $H_{0j}$ is initialized to be an identity matrix and $P_0$ is initialized to be an all-zero vector. In addition, the estimate for variance $\sigma_{kj}^2$ is given by

\begin{equation}\label{update2}
\begin{aligned}
\begin{cases}
&m_{kj}=\nu m_{(k-1)j}+\epsilon_{kj},\\
&n_{kj}=\nu n_{(k-1)j}+ 1,\\
&\bar{\epsilon}_{kj}=m_{kj}/n_{kj},\\
&u_{kj}=(\frac{m_{kj}-1}{m_{kj}})^2+(\frac{1}{m_{kj}})^2,\\
&v_{kj}=m_{kj}(1-u_{kj}),\\
&\sigma_{j}^2\leftarrow\frac{1}{v_{kj}}[(\nu v_{kj})\sigma_{j}^2+\frac{m_{kj}-1}{m_{kj}}(\bar{\epsilon}_{kj}-\epsilon_{kj})^2],
\end{cases}
\end{aligned}
\end{equation}
where $m_{kj}$ and $n_{kj}$ are initialized to be 0.

Eq. (\ref{lrm}) can characterize the spatial-temporal variation of charging demand. Different locations may have different charging demands. Thus, we use different linear regression equations to model these geographically separated charging stations. Furthermore, the price elasticity coefficients are updated continually using RLS algorithm defined in Eq. (\ref{update1}) and Eq. (\ref{update2}). The forgetting factor $\nu$ enables us to capture the most recent trend in charging demand and forget the outdated information. Thus, the RLS updating mechanism is able to track charging demand fluctuation over time.

\section{Impact on Power Grid from EV Charging}
For power flow analysis, we assume that an $N$-bus power network has 1 slack bus, $M$ load buses (PQ buses), and $N-M-1$ voltage-controlled buses (PV buses) \cite{sfrank}. Three phase balance operation and per-unit (p.u.) system are basic assumptions here. Charging stations are deployed across different PQ buses. Solving the power flow requires determining $N-1$ voltage phases (corresponding to PQ buses and PV buses) and $M$ voltage magnitude (corresponding to PQ buses). This is done by solving $N+M-1$ nonlinear power flow equations ($N-1$ active power equations and $M$ reactive power equations). The active and reactive power flow equations for each bus are given as follows,

\begin{equation}\label{networkflow1}
P_i=v_i\sum_{k=1}^Nv_k(G_{ik}\cos(\delta_{i}-\delta_{k})+B_{ik}\sin(\delta_{i}-\delta_{k})),
\end{equation}
\begin{equation}\label{networkflow2}
Q_i=v_i\sum_{k=1}^Nv_k(G_{ik}\sin(\delta_{i}-\delta_{k})-B_{ik}\cos(\delta_{i}-\delta_{k})),
\end{equation}
where $v_i$ and $\delta_i$ are, respectively, voltage magnitude and phase at the $i$-th bus; $P_i$ and $Q_i$ are real power and reactive power injections at the $i$-th bus; $G_{ik}$ and $B_{ik}$ are, respectively, conductance and susceptance of the $ik$-th element of the bus admittance matrix.

An increasing EV charging demand at PQ buses will lead to network-wide voltage variation (magnitude and phase) if the network does not provide sufficient active power and reactive power. We will use voltage variation as a metric to assess the impact of EV charging on power grid. Applying Newton's method, we can calculate the linear approximation of voltage variation in the following way

\begin{equation}
\begin{aligned}
\left[
    \begin{array}{c}
    \mathbf{\Delta V}\\
    \mathbf{\Delta \Phi}\\
    \end{array}
\right]&=\left[
    \begin{array}{cc}
    \mathbf{\frac{\partial P}{\partial V}} & \mathbf{\frac{\partial P}{\partial \Phi}}\\
    \mathbf{\frac{\partial Q}{\partial V}} & \mathbf{\frac{\partial Q}{\partial \Phi}}\\
    \end{array}
\right]^{-1}
\left[
    \begin{array}{c}
    \mathbf{\Delta P}\\
    \mathbf{\Delta Q}\\
    \end{array}
\right]\\
&=\mathbf{J}^{-1}
\left[
    \begin{array}{c}
    \mathbf{\Delta P}\\
    \mathbf{\Delta Q}\\
    \end{array}
\right],
\end{aligned}
\end{equation}
where $\mathbf{\Delta V}$ and $\mathbf{\Delta \Phi}$ are, respectively, vectors of magnitude variation and phase variation; $\mathbf{\Delta P}$ and $\mathbf{\Delta Q}$ are, respectively, vectors of increased active power and reactive power due to EV charging. In addition, $\mathbf{\frac{\partial P}{\partial V}}$ and $\mathbf{\frac{\partial P}{\partial \Phi}}$ are partial derivatives of active power with respect to voltage magnitudes and phases, and $\mathbf{\frac{\partial Q}{\partial V}}$, $\mathbf{\frac{\partial Q}{\partial \Phi}}$ are partial derivatives of reactive power with respect to voltage magnitudes and phases. In addition, $\mathbf{J}^{-1}$ is the inverse of Jacobian matrix from power flow equations, which is given by

\begin{equation}
\scalebox{0.87}{$
\mathbf{J}^{-1}=
\left[
    \begin{array}{cccc}
    b_{1,1} & b_{1,2} & \cdots & b_{1,N+M-1}\\
    b_{2,1} & b_{2,2} & \cdots & b_{2,N+M-1}\\
    \vdots & & & \vdots\\
    b_{N+M-1,1} & b_{N+M-1,2} & \cdots & b_{N+M-1,N+M-1}\\
    \end{array}
\right],$}
\end{equation}

Let the sequence $[a_1, a_2,\cdots,a_L]$ denote the bus indexes of all charging stations in the power network. For instance, $a_i(i=1,2\cdots,L)$ means that the $i$-th charging station is fed by the $a_i$-th bus in the power network.

Finally, we use the 2-norm voltage variation (magnitude and phase) to assess the impact of EV charging on power grid,

\begin{equation}
F_k=\left|\left|
    \mathbf{J}^{-1}
\left[
    \begin{array}{c}
    \mathbf{\Delta P}\\
    \mathbf{\Delta Q}\\
    \end{array}
\right]\right|\right|^2.
\end{equation}

Moreover, we denote $S_i^{\textrm{Ac}}$ and $S_i^{\textrm{Re}}$ as the active power sensitivity and reactive power sensitivity of the $i$-th PQ bus. And $S_i^{\textrm{Ac}}$ is defined as follows,

\begin{equation}\label{powersensitivity}
S_i^{\textrm{Ac}}=\left|\left|
    \mathbf{J}^{-1}
\left[
    \begin{array}{c}
    0\\
    \vdots\\
    0\\
    1\\
    0\\
    \vdots\\
    0
    \end{array}
\right]\right|\right|^2,
\end{equation}
where $S_i^{\textrm{Ac}}$ is the 2-norm voltage variation when the active power injection of the $i$-th PQ bus is increased by 1 W. Thus, 1 W is the $i$-th entry in the column vector in Eq. (\ref{powersensitivity}). Similarly, $S_i^{\textrm{Re}}$ is defined as the 2-norm voltage variation when the reactive power injection of the $i$-th PQ bus is increased by 1 var. A larger value of $S_i^{\textrm{Ac}}$ or $S_i^{\textrm{Re}}$ indicates that the PQ bus has a lower tolerance to load variation and more likely to disturb the whole network.

\vspace{-2mm}
\section{Stochastic Dynamic Programming (SDP) for Pricing and Electricity Procurement}
At first, this section introduces a safeguard of profit --- a minimum profit warning mechanism. In addition, major modules in SDP like renewable energy, real time wholesale electricity price, and system dynamics are discussed. Finally, we introduce the procedure to use SDP to derive pricing and electricity procurement policy.

\vspace{-1mm}
\subsection{A Safeguard of Profit}
In practice, service providers make decisions on pricing and electricity procurement based on the estimated charging demands. Although Eq. (\ref{lrm}) provides a viable way to estimate the charging demand, uncertainties still exist in actual charging demands. This subsection aims to develop a safeguard of profit to remind that the charging service provider should make a certain amount of profit under severe circumstance of uncertainties. We incorporate the safeguard as a constraint in the optimization framework. Wherever the optimal solution touches this constraint (i.e. this constraint becomes active), a warning will be raised for the service provider. The constraint is given as follows,

\begin{equation}\label{risk}
\textrm{Prob}\left(W_k<W_{\textrm{min}}\right)<\zeta,
\end{equation}
where $W_k$ is the profit made in the $k$-th horizon, $W_{\textrm{min}}$ is a profit threshold, and $\zeta$ is a small positive number in the range of $(0,1)$. Eq. (\ref{risk}) specifies that the probability that the actual profit is less than the profit threshold should be less than $\zeta$.

Expanding $W_k$ and rearranging terms in Eq. (\ref{risk}) yields the following

\begin{equation}
\textrm{Prob}\left(\mathbf{X}_k^{\textrm{T}}\mathbf{A}\mathbf{X}_k+\mathbf{B}^{\textrm{T}}\mathbf{X}_k+\mathbf{E}_k^{\textrm{T}}\mathbf{Z}_k+t_k<W_{\textrm{min}}\right)<\zeta,
\end{equation}
where $\mathbf{X}_k=[p_{k1},p_{k1},\cdots,p_{kL},o_k]^{\textrm{T}}$. Matrix A is given by
\begin{equation}
\mathbf{A}=
\left[
    \begin{array}{ccccc}
    -\gamma_{1,1} & \gamma_{1,2} & \cdots & \gamma_{1,L} & 0 \\
    \gamma_{2,1}  & -\gamma_{2,2} & \cdots & \gamma_{2,L} & 0 \\
    \vdots & \vdots & & \vdots &\vdots\\
    \gamma_{L,1} & \gamma_{L,2} & \cdots & -\gamma_{L,L} & 0 \\
    0 & 0 & \cdots& 0 & 0\\
    \end{array}
\right],
\end{equation}
and vector $\mathbf{B}$ is
\begin{equation}
\mathbf{B}=\left[\gamma_{0,1} + \frac{\eta_s}{\eta_d}\Gamma_1, \cdots, \gamma_{0,L} + \frac{\eta_s}{\eta_d}\Gamma_L, -c_k-\eta_s\eta_c\right]^{\textrm{T}},
\end{equation}
where $\Gamma_j$ is
\begin{equation}
\Gamma_j = -\gamma_{j,j} + \sum_{i = 1,i\neq j}^L\gamma_{j,i},
\end{equation}
and vector $E$ is
\begin{equation}
\mathbf{E}_k=\left[p_{k1} + \frac{\eta_s}{\eta_d}, \cdots,  p_{kL} + \frac{\eta_s}{\eta_d}, -\eta_s\right]^{\textrm{T}},
\end{equation}
and $\mathbf{Z}_k=[\epsilon_{k1},\epsilon_{k2},\cdots,\epsilon_{kL},w_k]^{\textrm{T}}$, and $t_k=\eta_s(\Gamma_0/\eta_d-I_k-\eta_cu_k)$, where $\Gamma_0$ is

\begin{equation}
\Gamma_0=\sum_{i=1}^L\gamma_{0,i}.
\end{equation}

Besides, we assume that $[\epsilon_{k1},\epsilon_{k2},\cdots,\epsilon_{kL},w_k]^{\textrm{T}}$ are independent Gaussian random variables. Thus, $\mathbf{E}_k^{\textrm{T}}\mathbf{Z}_k$ is also a Gaussian random variable with mean 0 and variance $\sum_{j=1}^L(p_{kj}+\eta_s/\eta_d)^2\sigma_{kj}^2+\eta_s^2\sigma_w^2$.

Finally, Eq. (\ref{risk}) can be rewritten as follows,
\begin{equation}\label{insurance}
\begin{aligned}
&\textrm{Prob}\left(\mathbf{E}_k^{\textrm{T}}\mathbf{Z}_k<W_{\textrm{min}}-\mathbf{X}_k^{\textrm{T}}\mathbf{A}\mathbf{X}_k-
\mathbf{B}^{\textrm{T}}\mathbf{X}_k-t_k\right)\\
&=\Phi\left(\frac{W_{\textrm{min}}-\mathbf{X}_k^{\textrm{T}}\mathbf{A}\mathbf{X}_k-
\mathbf{B}^{\textrm{T}}\mathbf{X}_k-t_k}{\sqrt{\sum_{j=1}^L(p_{kj}+\eta_s/\eta_d)^2\sigma_{kj}^2+\eta_s^2\sigma_w^2}}\right)
<\zeta,\\
\end{aligned}
\end{equation}
where $\Phi(.)$ is the cumulative distribution function (CDF) of a standard Gaussian random variable.

\subsection{Renewable Energy and Real Time Wholesale Price}
Literature abounds on various approaches to forecasting renewable energy, e.g., physical approach \cite{llandberg}-\cite{mlange}, statistical approach \cite{yzli}-\cite{eizgia}, and hybrid approach \cite{ggiebel}. In this paper, we use a Markov chain model \cite{jnorris}-\cite{pbremaud} which is a statistical approach, to demonstrate how renewable energy prediction is incorporated into our optimization model. In fact, other forecasting approaches can also be used in our model.

Markov chain characterizes the transition from the current renewable energy $u_k$ to the next $u_{k+1}$. We discretize renewable energy into $D$ levels, and the transition matrix at the $k$-th horizon is given by

\begin{equation}
\mathbf{T}_k=
\left[
    \begin{array}{cccc}
    t_{k,1,1} & t_{k,1,2} & \cdots & t_{k,1,D}\\
    t_{k,2,1} & t_{k,2,2} & \cdots & t_{k,2,D}\\
    \vdots & & & \vdots\\
    t_{k,D,1} & t_{k,D,2} & \cdots & t_{k,D,D}\\
    \end{array}
\right],
\end{equation}
where $t_{k,i,j}$ is the transition probability of renewable energy from level $i$ to level $j$ in the $k$-th horizon, and $\sum_{j=1}^Dt_{k,i,j}=1$. All transition probabilities can be estimated from historical data.

Similar to renewable energy, real time wholesale price forecasting has also been extensively studied through time series analysis, machine learning, big data, or hybrid approach in \cite{yji}-\cite{rweron}. Real time price forecasting is a topic beyond the technical scope of our paper. Thus, we do not study specific real time price forecasting approaches in this paper.

\subsection{Stochastic Dynamic Programming}
Eq. (\ref{eq1}) is a complex multi-variable optimization problem involving $K(L+1)$ variables. It may be mathematically cumbersome and difficult to solve in a brute-force manner. We observe that the original problem exhibits the properties of overlapping subproblems and optimal substructure, which can be solved efficiently using SDP. SDP solves a large-scale complex problem by partitioning it into a set of smaller and simpler subproblems \cite{gnemhauser}-\cite{dbertsekas}. The solution to the original problem is constructed by solving and combining the solutions of subproblems in a forward or backward manner. In contrast to a brute-force algorithm, SDP can greatly reduce computation and save storage.

In a wholesale real time electricity market, electricity is sold on an hourly basis. So our problem should have a finite number of planning horizons with $K=24$. System dynamics are governed by the evolution of system states, under the influence of decision variables and random variables. In our case, system dynamics are expressed by the following equations

\begin{equation}
\begin{aligned}
&I_{k+1}=I_{k}+\eta_cu_k+\eta_co_k-\frac{1}{\eta_d}\phi_k+w_k,\\
&u_{k+1}=h(u_k,\upsilon_k),\\
\end{aligned}
\end{equation}
where $I_k$ represents electricity storage at the beginning of the $k$-th horizon, $u_k$ is renewable energy, $o_k$ is the electricity procurement, $\eta_c$ is the charging efficiency, $\eta_d$ is the discharging efficiency, and $\phi_k$ is the total charging demand. Besides, $w_k$ and $\upsilon_k$ are independent process noises for the energy storage system and the renewable energy generation.

The aggregated expected utility from the first horizon to the $K$-th horizon is given by

\begin{equation}
\mathbb{E}\left\{\Pi_{K+1}(I_{K+1},u_{K+1})+\sum_{k=1}^K\Pi_k(I_k,u_k)\right\},
\end{equation}
where $\Pi_{K+1}(I_{K+1},u_{K+1})$ is a terminal utility occurred at the end of this process, and the expectation is taken over  $\epsilon_{kj}(j=1,\cdots,L)$ defined in Eq. (\ref{lrm}), $w_k$, and $\upsilon_k$. Therefore, the maximum aggregated expected utility $J(I_1,u_1)$ is given by

\begin{equation}\label{eq2}
\begin{aligned}
&J_1(I_1,u_1) =\max_{\mathbf{X}_1,
\cdots,\mathbf{X}_K}\mathbb{E}\left\{\Pi_{K+1}+\sum_{k=1}^K\Pi_k\right\},\\
&s.t.\\
&\begin{cases}
\textrm{Prob}(W_k<W_{\textrm{min}})<\zeta\\
0\leq o_k\leq o_{\textrm{max}};k=1,2,\cdots,N\\
p_{kj}\geq0;j=1,2,\cdots,L\\
 I_k + u_k + o_k-\sum_{j=1}^Ld_{kj}\geq0\\
I_k + u_k+ o_k-\sum_{j=1}^Ld_{kj}\leq E\\
d_{kj}\geq0;j=1,2,\cdots,L.
\end{cases}
\end{aligned}
\end{equation}
Applying SDP we can partition the problem into multiple small subproblems, which can be calculated recursively as follows,

\begin{equation}
\begin{aligned}
J_k(I_k,u_k) &=\max_{\mathbf{X}_k\in U_k(\mathbf{X}_k)}\mathbb{E}\left\{\Pi_k+J_{k+1}(I_{k+1},u_{k+1})\right\}\\
&=\max_{\mathbf{X}_k\in U_k(\mathbf{X}_k)}\left\{\mathbb{E}\{\Pi_k(I_k,u_k)\}+\right.\\
&\;\;\;\;\;\;\;\;\;\;\;\;\;\;\left.\mathbb{E}\{J_{k+1}(I_{k+1},u_{k+1})\}\right\}.
\end{aligned}
\end{equation}

Furthermore, we can rewrite each subproblem into a nice quadratic form by combining like terms as follows,
\begin{equation}
\begin{aligned}
J_k(I_k,u_k)=&\max_{\mathbf{X}_k\in U_k(\mathbf{X}_k)}\Big\{\mathbb{E}\{\frac{1}{2}\mathbf{X}_k^{\textrm{T}}\mathbf{Q}\mathbf{X}_k\\
&+\mathbf{B}^{\textrm{T}}_k\mathbf{X}_k\}+ \mathbb{E}\{r_k\}\Big\},
\end{aligned}
\end{equation}
where $\mathbf{Q}$, $\mathbf{B}_k$, and $r_k$ are given by

\begin{strip}
\vspace{1mm}
\hrule
\begin{equation}
\scalebox{0.95}{$
\mathbf{Q}=
\left[
    \begin{array}{cccc}
    -2\gamma_{1,1}\lambda_1 - \alpha\lambda_2\Gamma_1^2-2\lambda_3\sum_{j=1}^{N+M-1}\Theta_{1,j}^2 & \cdots & 2\gamma_{1,L}\lambda_1 - \alpha\lambda_2\Gamma_1\Gamma_L-2\lambda_3\sum_{j=1}^{N+M-1}\Theta_{1,j}\Theta_{L,j}&0\\
    2\gamma_{2,1}\lambda_1-\alpha\lambda_2\Gamma_2\Gamma_1 - 2\lambda_3\sum_{j=1}^{N+M-1}\Theta_{2,j}\Theta_{1,j}& \cdots & 2\gamma_{2, L}\lambda_1 -\alpha\lambda_2\Gamma_2\Gamma_L-2\lambda_3\sum_{j = 1} ^ {N+M-1}\Theta_{2,j}\Theta_{L,j}& 0 \\
    \vdots & & & \vdots\\
    2\gamma_{L,1}\lambda_1 - \alpha\lambda_2\Gamma_L\Gamma_1-2\lambda_3\sum_{j=1}^{N+M-1}\Theta_{L,j}\Theta_{1,j}& \cdots & -2\gamma_{L,L}\lambda_1 - \alpha\lambda_2\Gamma_L^2-2\lambda_3\sum_{j=1}^{N+M-1}\Theta_{L,j}^2& 0\\
    0 & \cdots & 0 & 0\\
    \end{array}
\right],$}
\end{equation}
\end{strip}

\begin{strip}
\begin{equation}
\mathbf{B}_k =
\left[
    \begin{array}{c}
    \lambda_1(\gamma_{0,1}+\frac{\eta_s}{\eta_d}\Gamma_1) + \lambda_2\left(\omega\sum_{j=1}^L\gamma_{1,j}-\alpha\Gamma_0\Gamma_1\right)-
    2\lambda_3\sum_{j=1}^{N+M-1}\Theta_{0,j}\Theta_{1,j}\\
    \vdots\\
    \lambda_1(\gamma_{0,L}+\frac{\eta_s}{\eta_d}\Gamma_L)+ \lambda_2\left(\omega\sum_{j=1}^L\gamma_{L,j}-\alpha\Gamma_0\Gamma_L\right)-
    2\lambda_3\sum_{j=1}^{N+M-1}\Theta_{0,j}\Theta_{L,j}\\
    -(c_k+\eta_s\eta_c)\lambda_1\\
    \end{array}
\right],
\end{equation}
\end{strip}

\begin{strip}
\begin{equation}
\begin{aligned}
r_k=&\lambda_1\eta_s(\Gamma_0/\eta_d-I_k-\eta_cu_k)+\lambda_2\left(\omega\Gamma_0-\frac{\alpha}{2}\left(\Gamma_0^2+\sum_{j=1}^L\sigma_{k,j}^2\right)\right)-\\
&\lambda_3\left(\sum_{j=1}^{N+M-1}\Theta_{0,j}^2+\lambda_3\sum_{j=1}^{N+M-1}\sum_{i=1}^Lb_{j,a_i}^2\sigma_{k,j}^2\right)
+\mathbb{E}_{u_{k+1}}\{J_{k+1}(I_{k+1},u_{k+1})\},
\end{aligned}
\end{equation}

\hrulefill
\end{strip}

\noindent where $\Theta_{n,j}$ is

\begin{equation}
\Theta_{n,j}=-b_{j,a_n}\gamma_{n,n}+\sum_{i=1,i\neq n}^Lb_{j,a_i}\gamma_{i,n},
\end{equation}
and $a_n$ is the bus index of the power network for the $n$-th charging station, and

\begin{equation}
\Theta_{0,j}=\sum_{i=1}^Lb_{j,a_i}\gamma_{0,i}.
\end{equation}

\begin{figure}[htbp]
\centering
\includegraphics[width=3in]{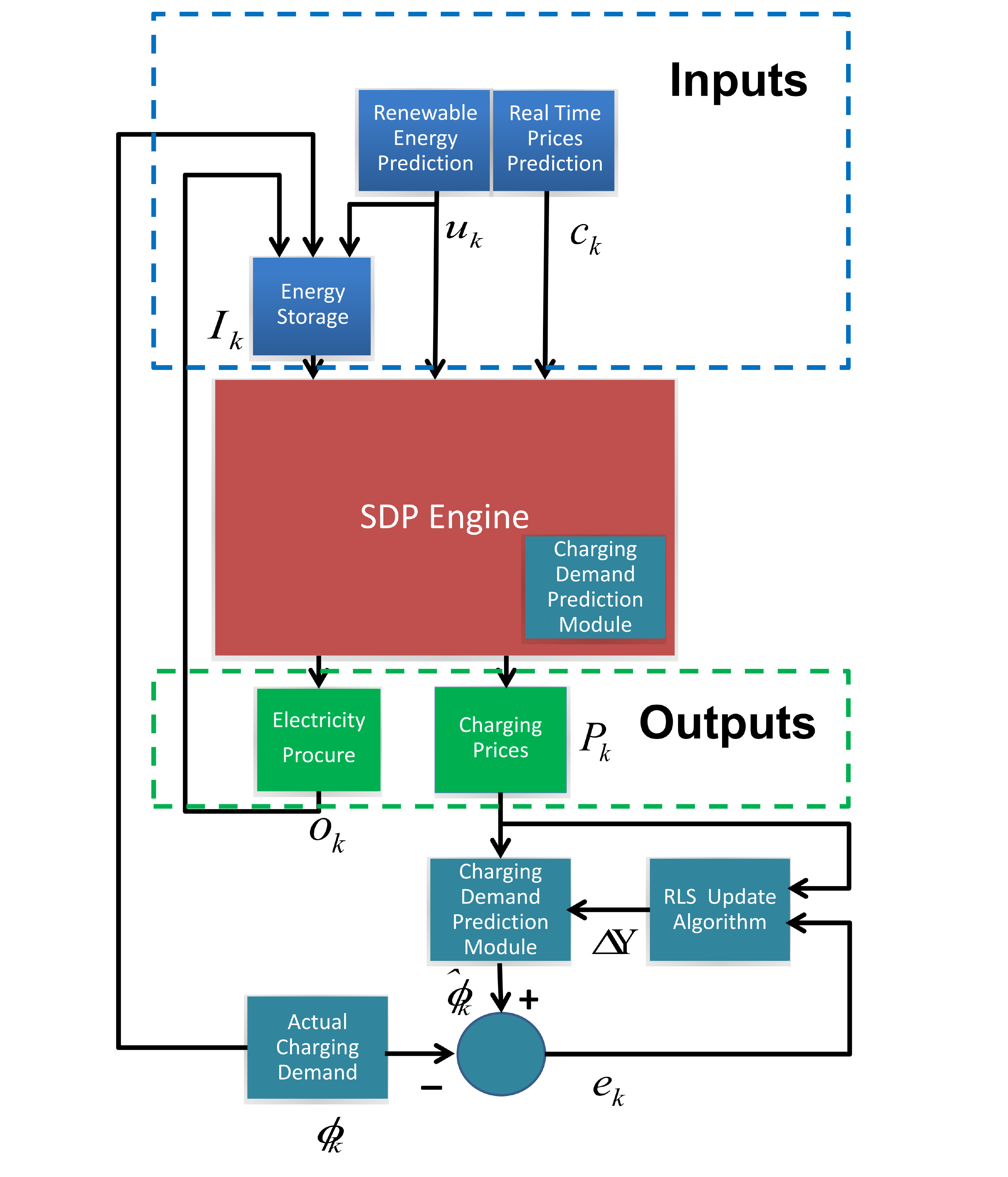}
\caption{Dynamic Pricing and Energy Management Algorithm}
\label{fig3}
\end{figure}

Finally, $J_{k+1}(I_{k+1},u_{k+1})$ is the total aggregated utility starting from the $(k+1)$-th horizon to the $K$-th horizon. Fig. \ref{fig3} illustrates the schematic of the entire optimization framework. The charging service provider should run the SDP engine at the beginning of every planning horizon.

\section{Simulations and Discussions}
\begin{figure*}[htbp]
  \begin{minipage}[b]{0.5\linewidth}
    \includegraphics[width=3in]{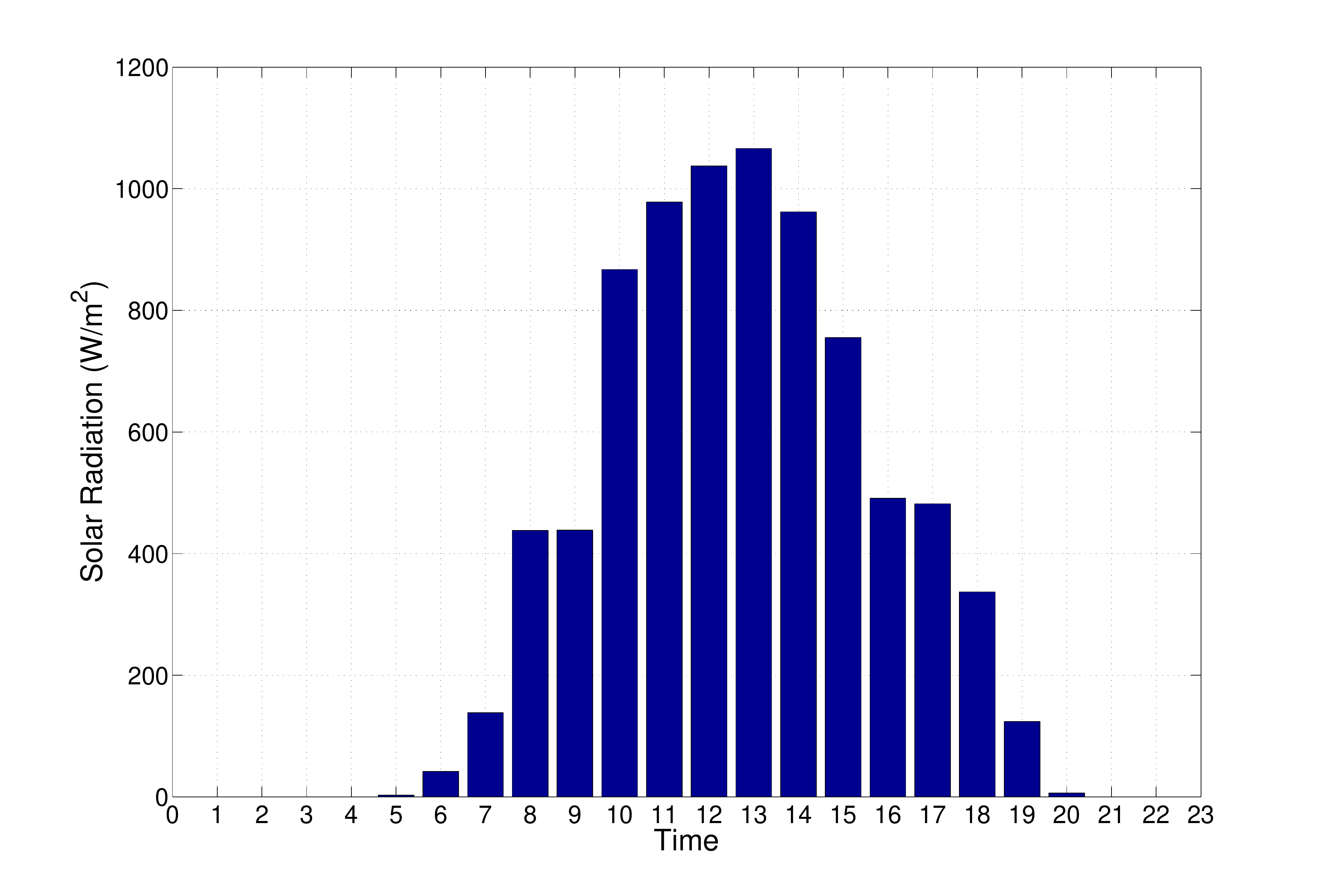}
    \caption{Typical Daily Solar Radiation}
    \label{fig5}
  \end{minipage}
    \begin{minipage}[b]{0.5\linewidth}
    \includegraphics[width=3in]{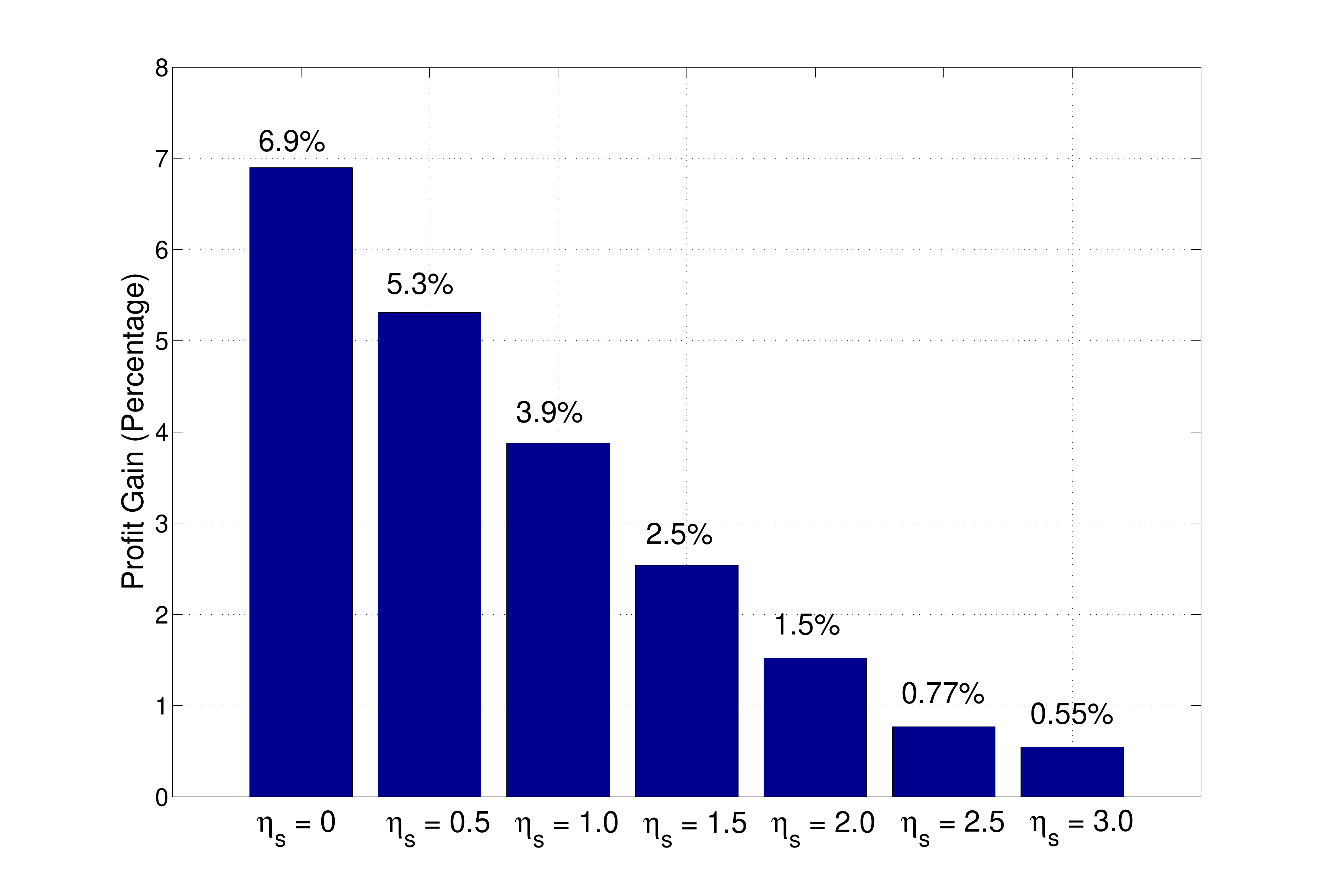}
    \caption{SDP Profit Increase Percentage}
    \label{dp_vs_greedy}
  \end{minipage}
  \end{figure*}

The simulation coefficients are given in Table \rom{1}. Tesla's home rechargeable Lithium-ion battery system --- Powerwall has a 92.5\% round-trip DC efficiency with 100\% depth of discharge \cite{powerwall}. Eos Energy Storage has a battery-based energy storage with a round-trip efficiency of 75\% and a 100\% depth of discharge \cite{esoenergy}. In our simulation, we assume the charging efficiency $\eta_c$ and discharging efficiency $\eta_d$ are both 0.9. For simplicity, we use the day-ahead wholesale electricity price data from PJM \cite{pjmcom} to represent the real time wholesale price forecasting in the simulations, but other forecasting approaches can be used. In addition, we assume that the charging service provider procures electricity at a single locational marginal price (LMP). We use solar power to represent the renewable energy source. The solar radiation data is from National Renewable Energy Laboratory (NREL) \cite{emckenna}, and the typical daily solar radiation is depicted in Fig. \ref{fig5}. Note that solar radiation begins at 6:00 am and ends at 8:00 pm. Additionally, we assume that solar cell efficiency is 20\%. We use IEEE 57 Bus Test case for the power network in our simulations \cite{testcase}.

\begin{table*}[!htbp]\label{ta1}
\caption{Simulation Parameters}
\begin{center}
\begin{tabular}{llll}
\hline
Coefficient & Description & Unit & Value\\

$N$ & Number of horizons & - &24\\

$E$ & Energy storage capacity & MWh & 200\\

$\lambda_1$ & Weight for profit & - & 0 to 1\\

$\lambda_2$ & Weight for customer satisfaction & - & 0 to 1\\

$\lambda_3$ & Weight for impact & - & 0 to 1\\

$\zeta$ & Revenue safeguard probability & - & 0.2\\

$\alpha$ & Shape parameter &-&5e-5\\

$\omega$ & Shape parameter &-&0.01\\

$\eta_s$ & Unit storage cost  & \$/MWh &0 to 4\\

$\eta_c$ & Energy storage charging efficiency & - & 0.9\\

$\eta_d$ & Energy storage discharging efficiency & - & 0.9\\

$\rho_0$ & Knee point threshold & - & 1\\
\hline
\end{tabular}
\end{center}
\end{table*}

\begin{figure*}[htbp]
  \begin{minipage}[b]{0.5\linewidth}
    \includegraphics[width=3in]{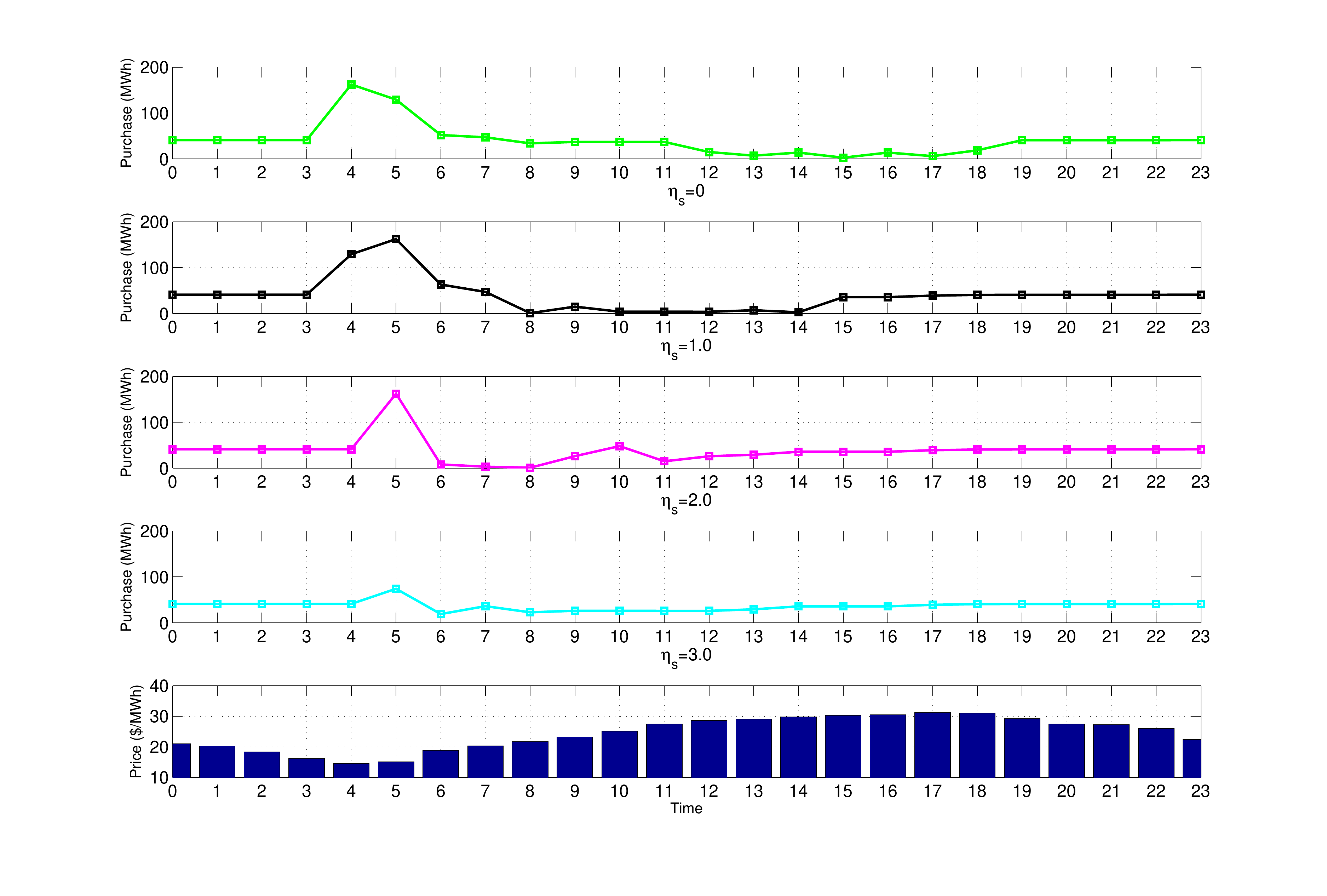}
    \caption{Electricity Procurement with Different Storage Cost}
    \label{purchase_vs_storage}
    \end{minipage}
      \begin{minipage}[b]{0.5\linewidth}
    \includegraphics[width=3in]{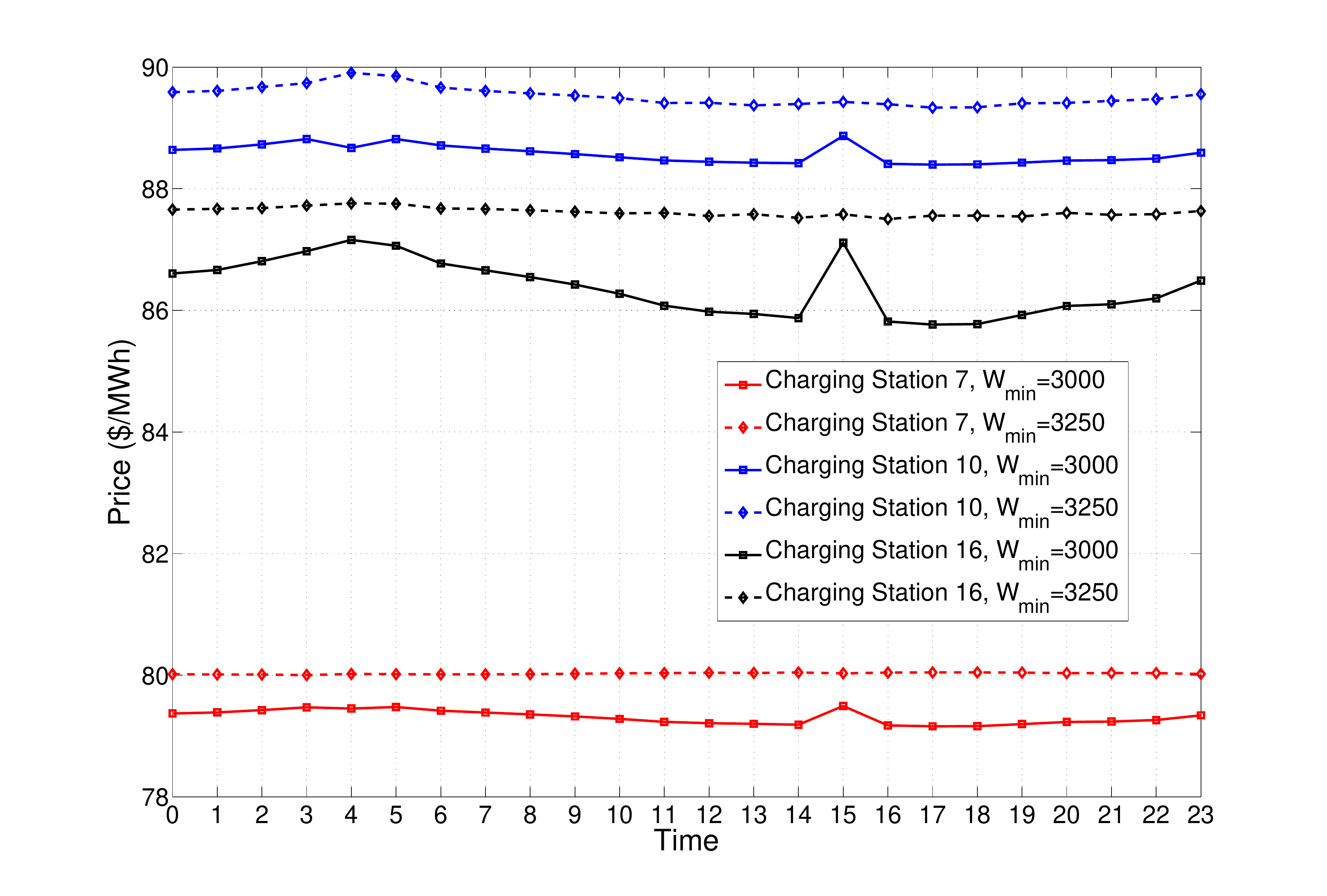}
    \caption{Charging Prices with Different $W_{\textrm{min}}$}
    \label{price_vs_rmin}
  \end{minipage}
  \end{figure*}

  \begin{figure*}[htbp]
  \begin{minipage}[b]{0.5\linewidth}
    \includegraphics[width=3in]{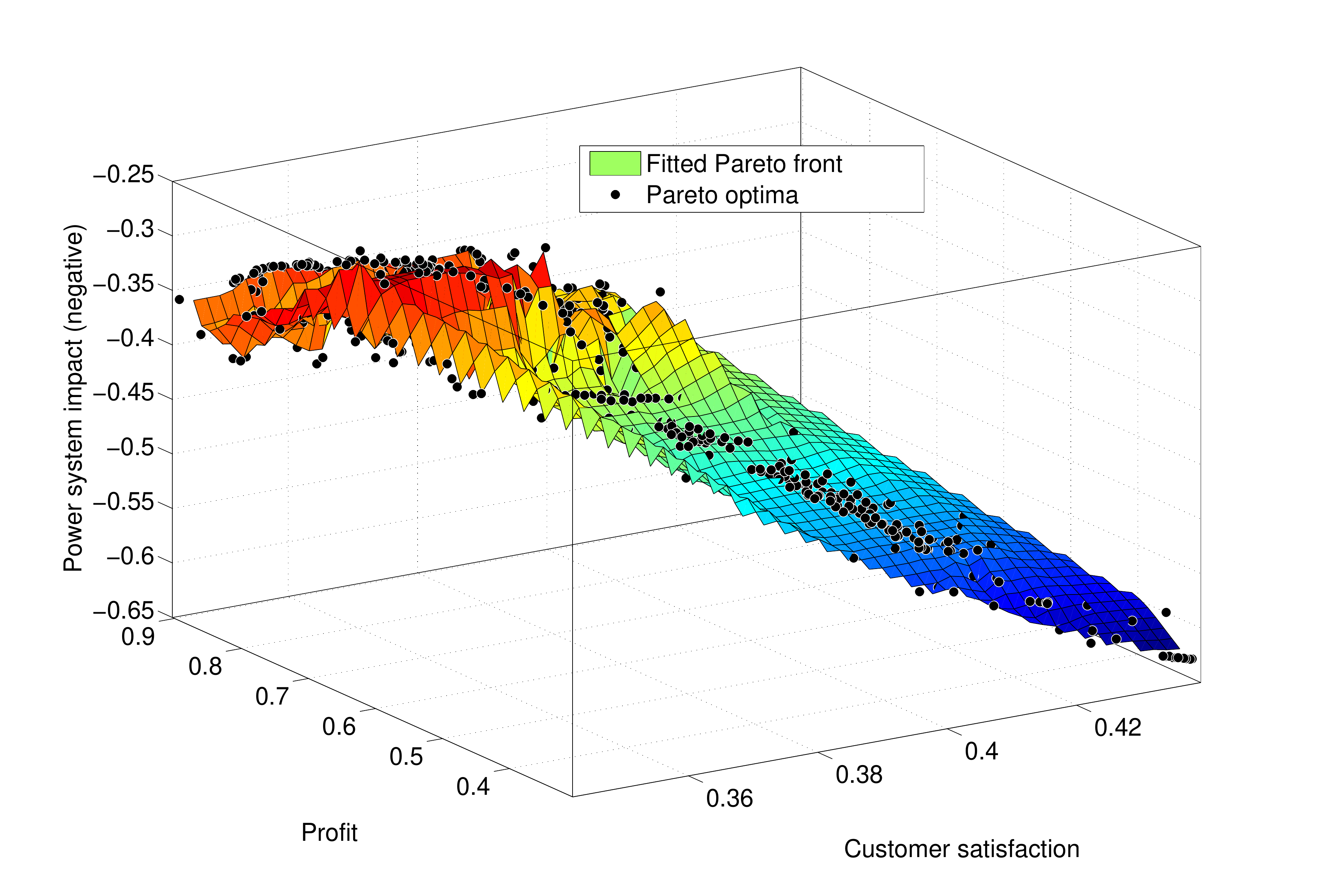}
    \caption{Pareto Front}
    \label{pareto_front}
  \end{minipage}
    \begin{minipage}[b]{0.5\linewidth}
    \includegraphics[width=3in]{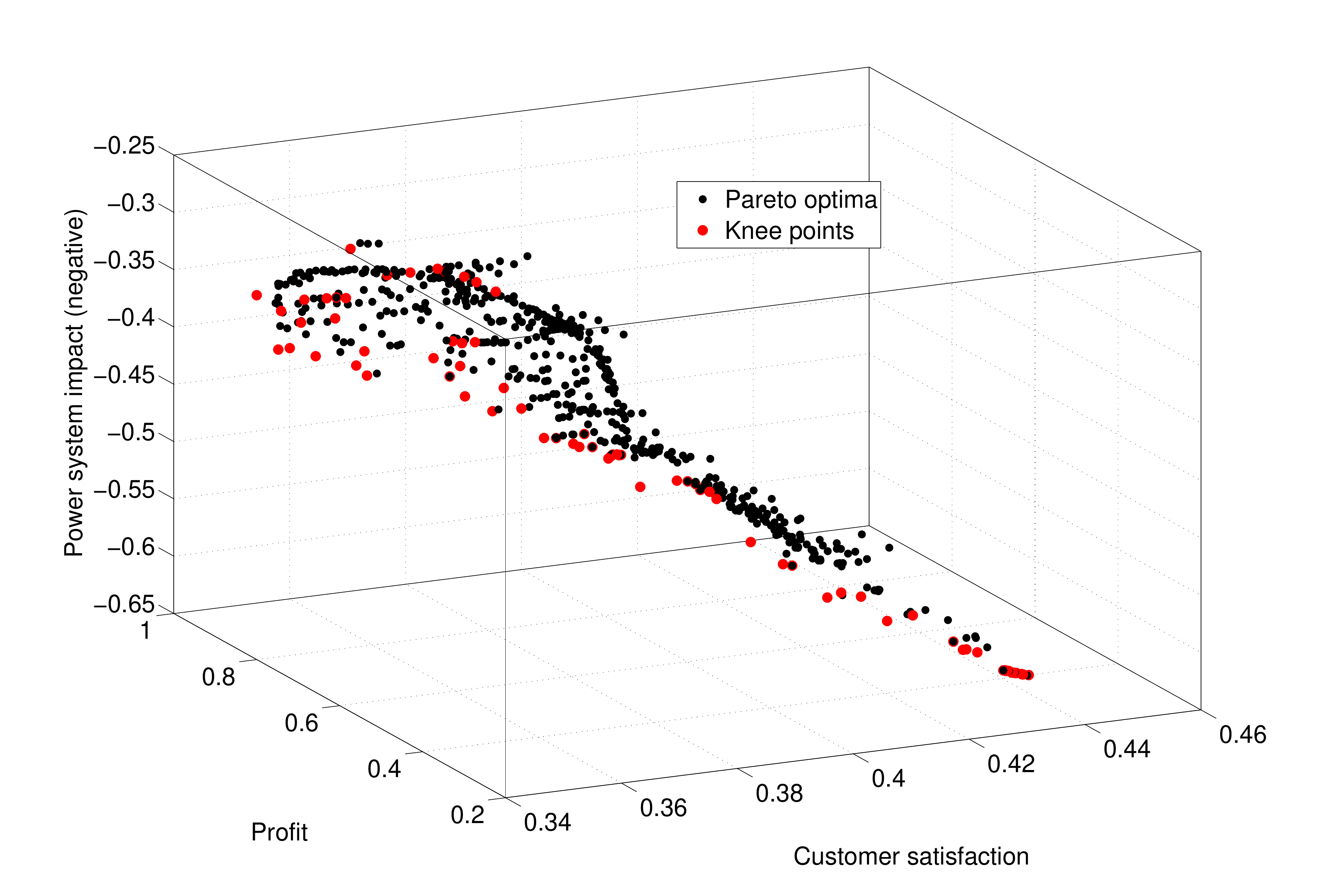}
    \caption{Pareto Front with Knee Points}
    \label{pareto_front_knee_points}
  \end{minipage}
\end{figure*}

\begin{figure*}[htbp]
  \begin{minipage}[b]{0.5\linewidth}
    \includegraphics[width=3in]{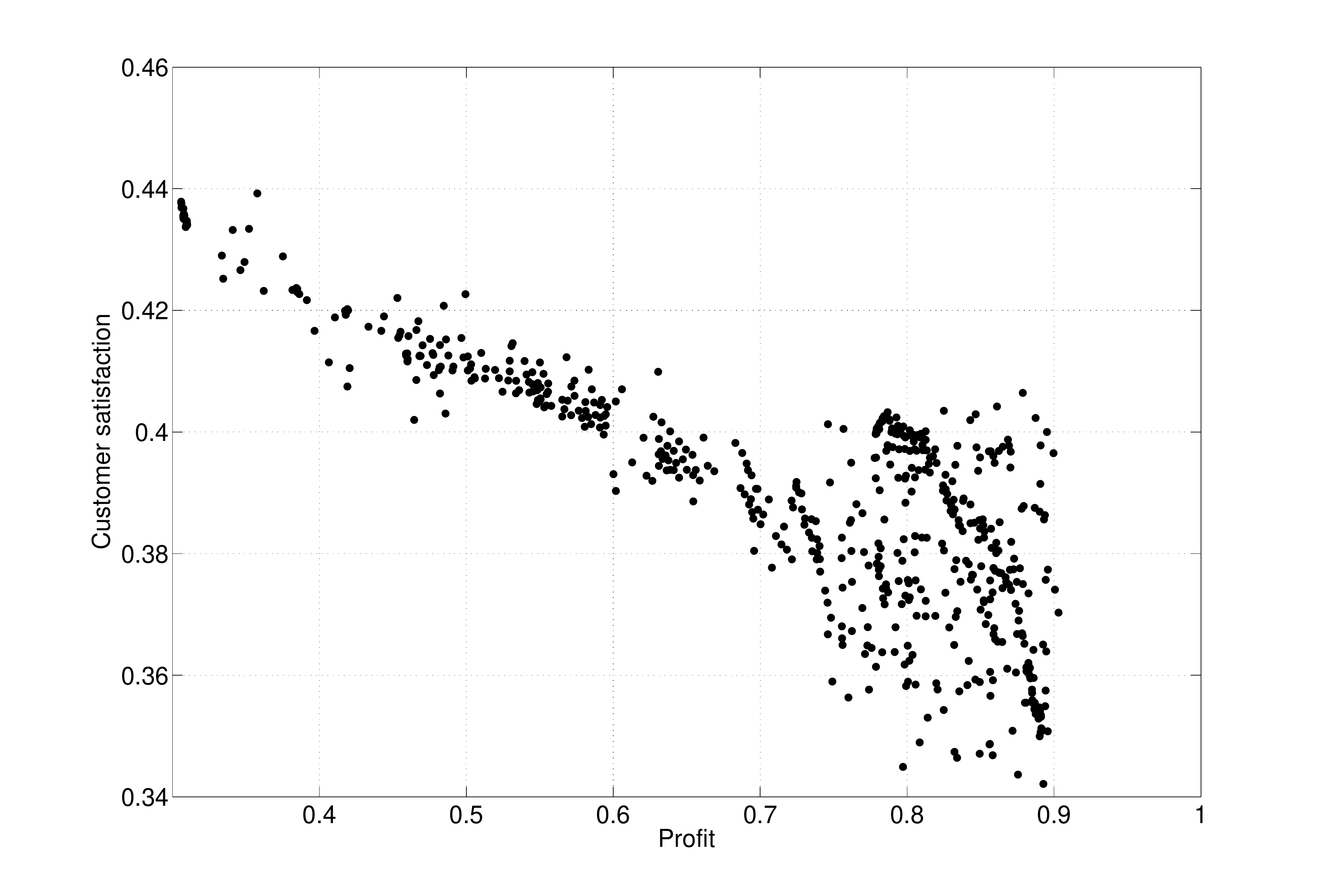}
    \caption{Profit vs Customer Satisfaction}
    \label{profit_vs_customer}
  \end{minipage}
  \begin{minipage}[b]{0.5\linewidth}
    \includegraphics[width=3in]{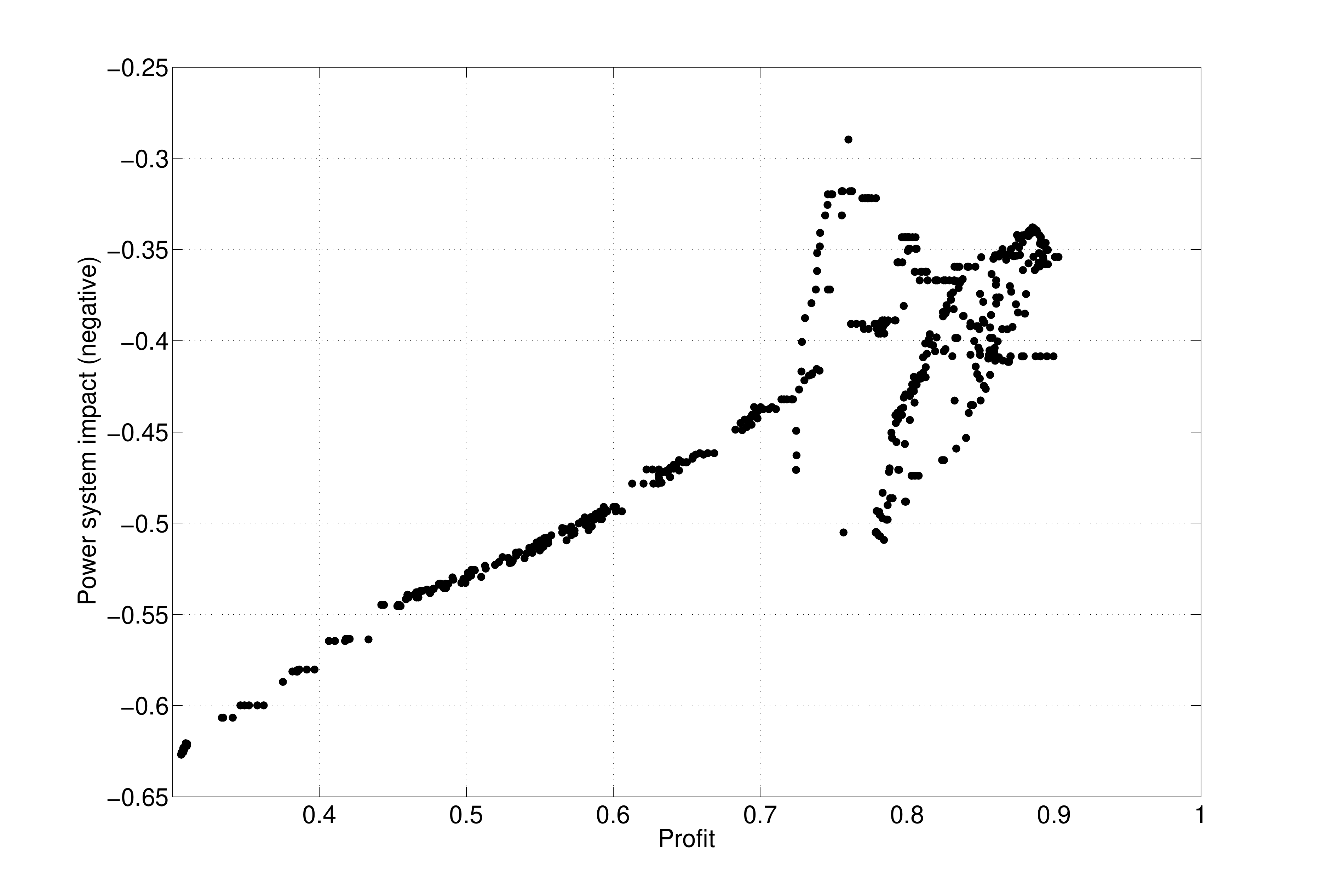}
    \caption{Profit vs Impact on Power Grid}
    \label{profit_vs_impact}
  \end{minipage}
\end{figure*}

\begin{figure*}[htbp]
  \begin{minipage}[b]{0.5\linewidth}
    \includegraphics[width=3in]{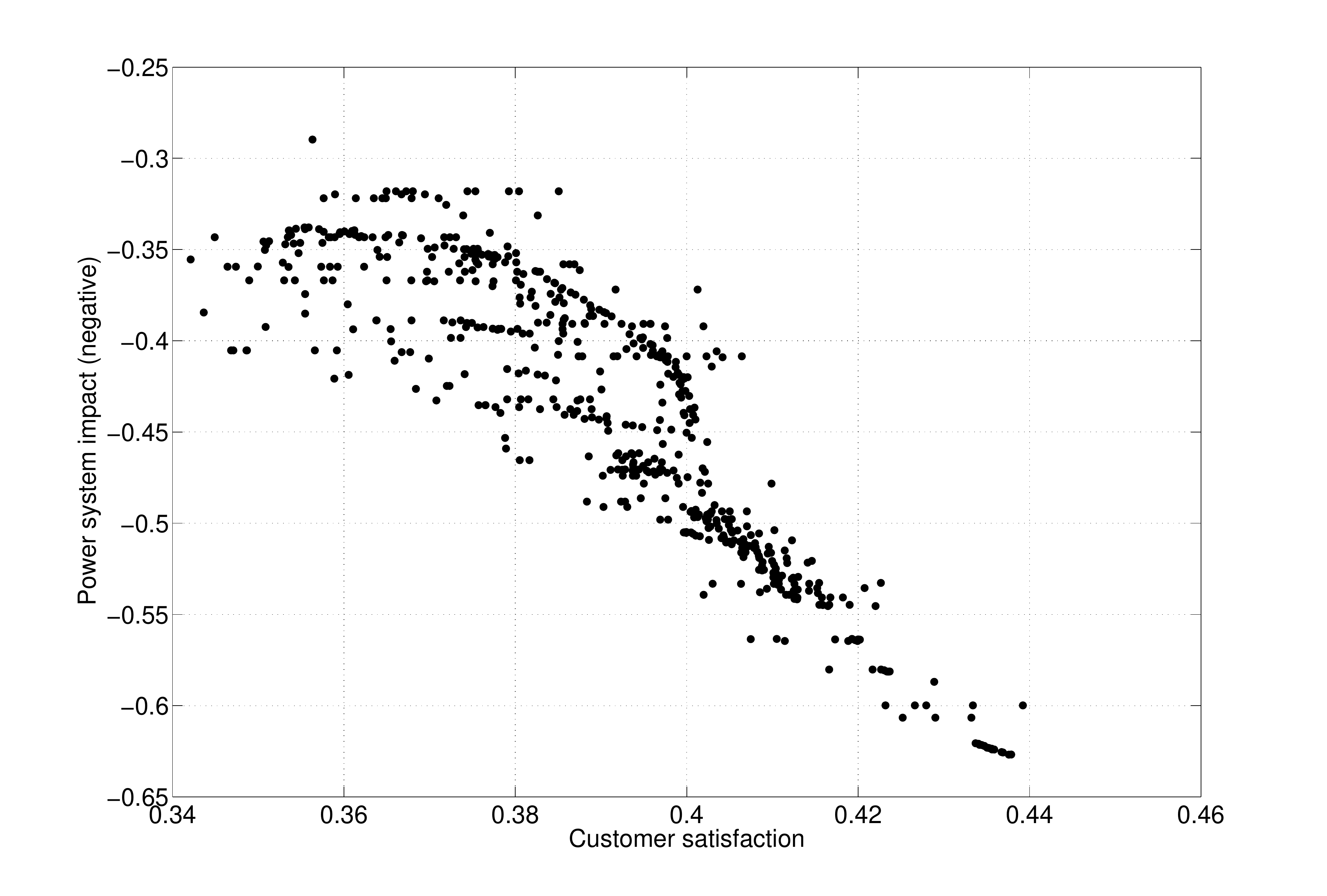}
    \caption{Customer Satisfaction vs Impact on Power Grid}
    \label{customer_vs_impact}
  \end{minipage}
  \begin{minipage}[b]{0.5\linewidth}
    \includegraphics[width=3in]{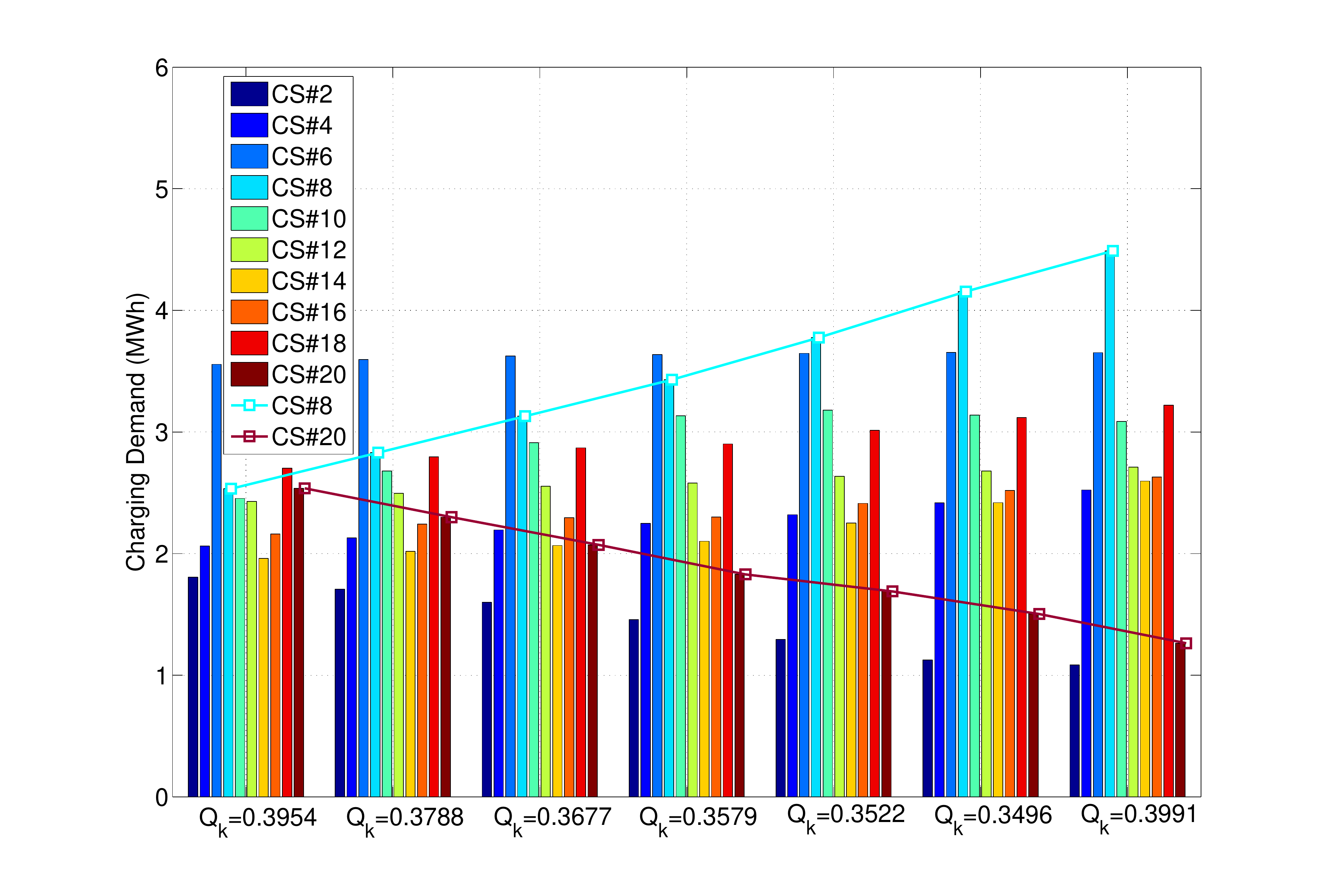}
    \caption{Charging Demand Redistribution with Different Impacts}
    \label{charging_vs_parameter}
  \end{minipage}
\end{figure*}

\subsection{SDP Algorithm versus Greedy Algorithm}
The greedy algorithm aims to optimize the current planning horizon without considering the future. We use the greedy algorithm as a benchmark, to which we compare the SDP algorithm in terms of profitability. The profit percentage gain of SDP algorithm compared to greedy algorithm is shown in Fig. \ref{dp_vs_greedy}. The simulation reveals that the SDP algorithm can achieve up to 7\% profit gain compared to the greedy algorithm. The reason why SDP is able to obtain a higher profit is that it fully exploits the information of day-ahead wholesale electricity prices and renewable energy prediction, and makes decisions to optimize the aggregated utility over multiple horizons. However, the greedy algorithm lacks a forward-looking vision, which solely maximizes the utility of the current horizon. As far as the computational complexity is concerned, greedy algorithm has a linear time complexity with $O(K)$, and SDP has a quadratic time complexity with $O(K^2)$, where $K$ is the number of planning horizons. This is because the greedy algorithm only involves one loop from horizon 0 to horizon 23. However, the SDP algorithm has two loops with the outer loop starting from horizon 0 to horizon 23 and the inner loop for backward recursive SDP calculation. In essence, the SDP algorithm trades complexity for a higher profit.

\subsection{Aggressive or Conservative Electricity Procurement Strategy}
An electricity storage enables the charging service provider to store the intermittent renewable energy or excessive electricity when the wholesale price is low, and sell it to EVs when the wholesale price is high. In this subsection, we analyse how this ``buy low and sell high" strategy will change when the unit storage cost ($\eta_s=0$ to 4) changes. In Fig. \ref{purchase_vs_storage}, electricity procurement strategies with different unit storage costs are depicted in the first four subplots, and the last subplot shows the real time wholesale electricity prices. We can make three observations: (1) From 8:00 to 16:00, the service provider tends to procure less electricity from the wholesale market because of renewable energy generation at this period of time, and (2) The service provider tends to procure more electricity during the low wholesale price period (from 3:00 to 6:00) and procure less electricity during the high wholesale price period (from 11:00 to 17:00), and (3) When $\eta_s$ is small, the service provider becomes aggressive in electricity procurement during low price period, and when $\eta_s$ is large, it becomes more conservative.

\subsection{Charging Price with Safeguard of Profit}
In the simulation, we investigate the interplay between charging prices and the safeguard of profit. From Fig. \ref{price_vs_rmin}, we note that the charging prices increase as the profit threshold $W_{\textrm{min}}$ increases. According to Eq. (\ref{insurance}), we must ensure the probability $\zeta$ does not change even if $W_{\textrm{min}}$ increases. In other words, $(W_{\textrm{min}}-\mathbf{X}_k^{\textrm{T}}\mathbf{A}\mathbf{X}_k-
\mathbf{B}^{\textrm{T}}\mathbf{X}_k-t_k)/(\sqrt{\sum_{j=1}^L(p_{kj}+\eta_s/\eta_d)^2\sigma_{kj}^2+\eta_s^2\sigma_w^2})$ should not change as $W_{\textrm{min}}$ increases. The simulation results show that the charging service provider ends up raising charging prices to ensure the probability $\zeta$.

\subsection{Pareto Optima and Knee Points}
We need to simultaneously maximize multiple objectives --- profit, customer satisfaction, and the negative of impact on power grid. Each point in Fig. \ref{pareto_front} is a Pareto optimum in which it is impossible to increase any one individual objective without decreasing at least one of the other objectives \cite{clhwang}. The Pareto front is obtained by using the linear interpolation fitting method \cite{pdavis}.

Knee points in the Pareto front provide the best tradeoff among multiple objectives, which yield largest improvement per unit degradation. Following the metric discussed in \cite{lrachmawati}-\cite{xzhang}, we define $\rho(Y_i,S)$ to represent the least improvement per unit degradation by replacing any other Pareto optima in $S$ with $Y_i$. The entries in $Y_i=[y_{1i},y_{2i},y_{3i}]^{\textrm{T}}$ represent the profit, the customer satisfaction, and the impact on power grid, respectively.

\begin{equation}
\rho(Y_i,S)=\min_{Y_j\in S, j\neq i}\frac{\sum_{k=1}^3\max(0,y_{ki}-y_{kj})}{\sum_{k=1}^3\max(0,y_{kj}-y_{ki})}.
\end{equation}
Then we set a threshold $\rho_0$ to select the knee points as follows,

\begin{equation}
S_{\textrm{knee}}^{\rho_0}=\{Y_i|\rho(Y_i,S)>\rho_0;Y_i\in S\}.
\end{equation}
We use $\rho=1$ in the simulations. The knee points are marked in red in Fig. \ref{pareto_front_knee_points}. We notice that there are several knee regions among the Pareto optima, which reflect different preference over the three objectives --- profit, customer satisfaction, and the impact on power grid.

\subsection{Interplays between Profit, Customer Satisfaction, and Impact on Power Grid}
The projection of Pareto optima on the Profit-Customer plane is plotted in Fig. \ref{profit_vs_customer}. We observe that customer satisfaction decreases when profit increases. This is because the charging service provider raises charging prices to decrease the total charging demand. The decreased total charging demand leads to a decreased customer satisfaction. However, the net effect of raising charging prices is that the service provider achieves a higher profit. Therefore, the service provider should strike a balance between the two competing objectives of profit and customer satisfaction.

The projection of Pareto optima on the Profit-Impact plane is plotted in Fig. \ref{profit_vs_impact}. It turns out that the impact and the profit are not competing objectives since the impact on power grid decreases when profit increases. The increased charging prices cause a decrease in total charging demand, relieving the stress on power grid. However, the profit is improved even though the total charging demand decreases.

Fig. \ref{customer_vs_impact} shows the projection of Pareto optima on the Customer-Impact plane. Note that customer satisfaction and the impact are competing objectives since the impact on power grid increases as customer satisfaction increases. It is obvious that customer satisfaction and impact on power grid are both related to the total charging demand. According to Eq. (\ref{satisfactioneq}), customer satisfaction increases when the total charging demand increases. However, the increased charging demand will inevitably pose a heavier stress on the power grid.

\subsection{Spatial Charging Demand versus Impact on Power Grid}
The relationship between spatial charging demand and impact on power grid is shown in Fig. \ref{charging_vs_parameter}. Due to limited space, we only plotted the charging stations with even indices. We observe that as the impact on power grid ($Q_k$) decreases, the charging demands of Charging Station \#2 (CS\#2) and Charging Station \#20 (CS\#20) decrease while the charging demands of other charging stations increase. This is because the PQ buses feeding CS\#2 and CS\#20 have larger active power sensitivity metric $S_i^{\textrm{Ac}}$ than the others. The active power sensitivity for charging stations with even indices are $[0.80, 0.61, 0.33, 0.17, 0.31, 0.29, 0.22, 0.60, 0.29, 1.33]$. Note that CS\#8 has the smallest active power sensitivity 0.17, its charging demand increases very fast as the impact decreases. While CS\#20 has the largest active power sensitivity 1.33, its charging demand decreases fast. Thus, the service provider has to shift the charging demands from the PQ buses with large $S_i^{\textrm{Ac}}$ to those with small $S_i^{\textrm{Ac}}$ to reduce the impact on power grid.

\section{Conclusion}
This paper proposes a multi-objective optimization framework for EV charging service provider to determine retail charging prices and appropriate amount of electricity to purchase from the real time wholesale market. A linear regression model is employed to estimate EV charging demands. To cope with multiple uncertainties, SDP algorithm is applied to simplify the optimization problem. Compared to greedy algorithm (benchmark), SDP algorithm can make a higher profit at the cost of increased algorithm complexity. A lost-cost electricity storage is beneficial for the service provider to harvest the intermittent renewable energy and exert the ``buy low and sell high" strategy to improve profits. In addition, the service provider can shift charging demands from high-sensitive buses to low-sensitive buses to alleviate the impact on power grid by changing charging prices.



\begin{IEEEbiography}[{\includegraphics[width=1.1in,height=1.2in,clip,keepaspectratio]{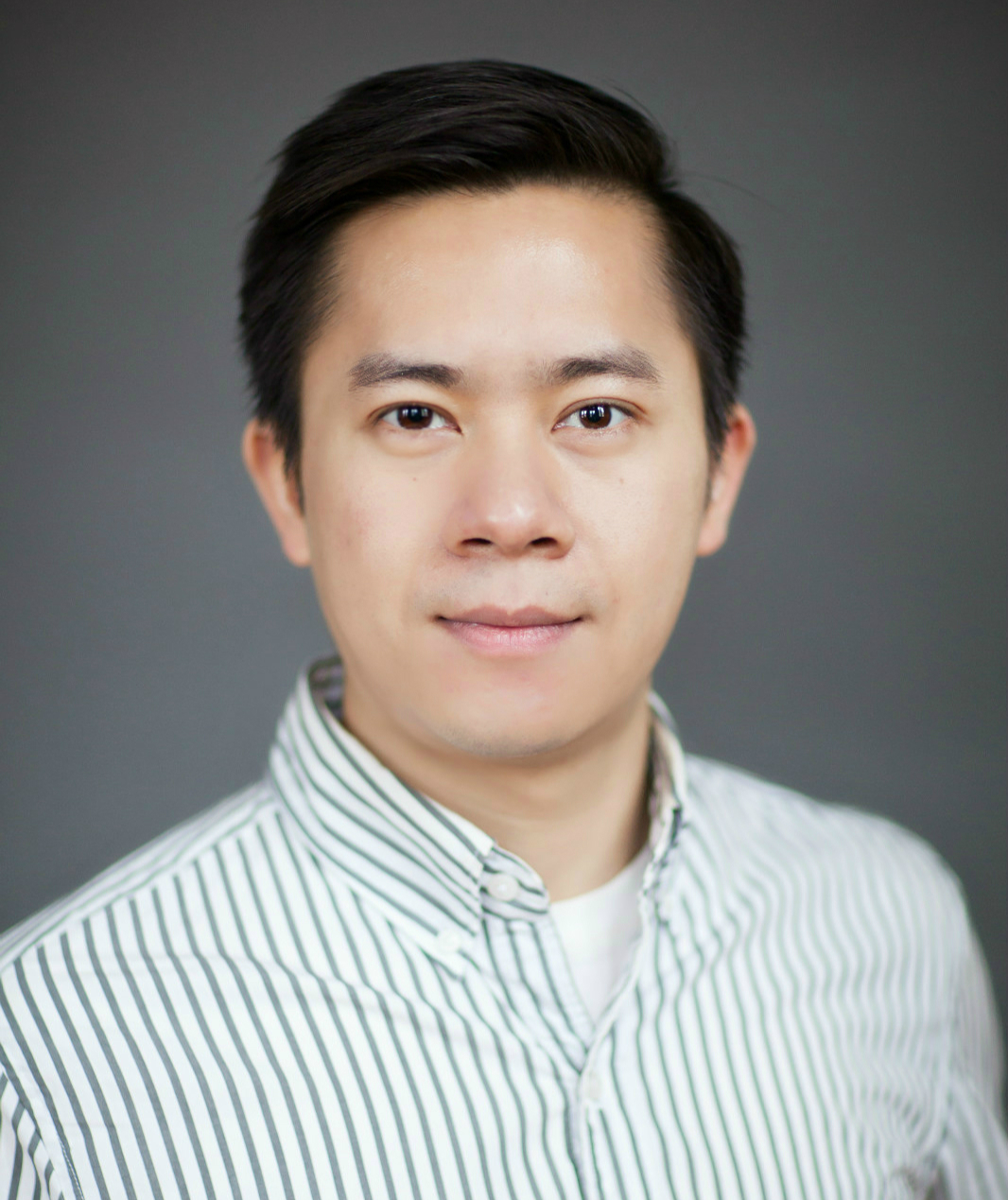}}]{Chao Luo}
received the B.Eng degree with distinction in Communication Engineering from Harbin Institute of Technology (HIT), China, in 2012. Currently, he is pursuing his Ph.D. in Electrical Engineering in the University of Notre Dame, USA. Chao's research interests include electric vehicle (EV) integration into power system, machine learning in smart grid, power system architecture, and electricity market.
\end{IEEEbiography}

\begin{IEEEbiography}[{\includegraphics[width=1in,height=1.25in,clip,keepaspectratio]{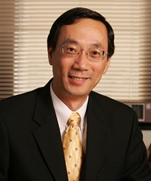}}]{Yih-Fang Huang}
is Professor of Department of Electrical Engineering and Senior Associate Dean of College of Engineering at University of Notre Dame.  Dr. Huang received his BSEE degree from National Taiwan University, MSEE degree from University of Notre Dame and Ph.D. from Princeton University. He served as chair of the Electrical Engineering Department at the University of Notre Dame from 1998 to 2006.  His research interests focus on theory and applications of statistical signal detection and estimation, and adaptive signal processing.

In Spring 1993, Dr. Huang received the Toshiba Fellowship and was Toshiba Visiting Professor at Waseda University, Tokyo, Japan.  From April to July 2007, he was a visiting professor at the Munich University of Technology, Germany.  In Fall, 2007, Dr. Huang was awarded the Fulbright-Nokia scholarship for lectures/research at Helsinki University of Technology in Finland (which is now Aalto University).

Dr. Huang received the Golden Jubilee Medal of the IEEE Circuits and Systems Society in 1999, served as Vice President in 1997-98 and was a Distinguished Lecturer for the same society in 2000-2001.  At the University of Notre Dame, he received Presidential Award in 2003, the Electrical Engineering department's Outstanding Teacher Award in 1994 and in 2011, the Rev. Edmund P. Joyce, CSC Award for Excellence in Undergraduate Teaching in 2011, and the College of Engineering's Teacher of the Year Award in 2013.  Dr. Huang is a Fellow of the IEEE.
\end{IEEEbiography}

\begin{IEEEbiography}[{\includegraphics[width=1in,height=1.25in,clip,keepaspectratio]{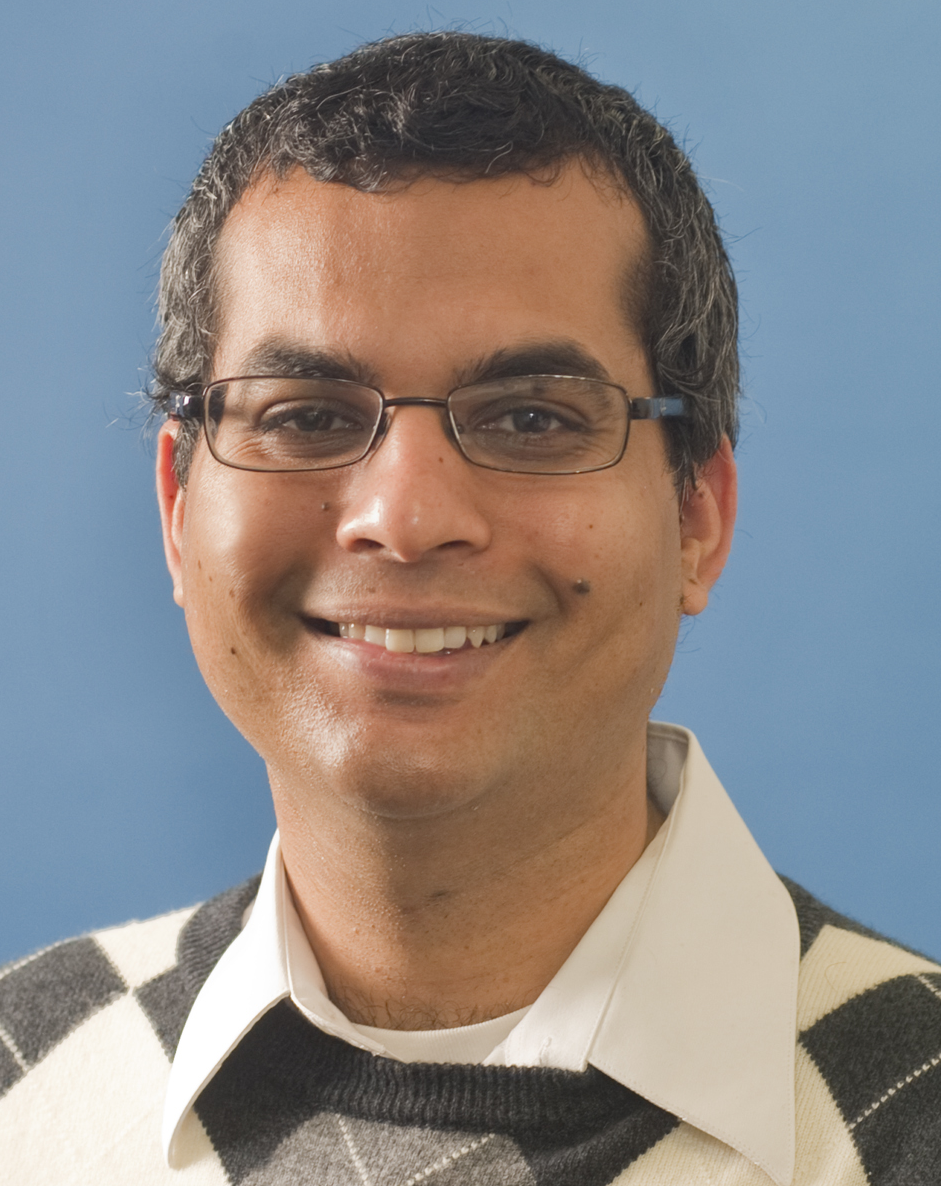}}]{Vijay Gupta}
is a Professor in the Department of Electrical Engineering at the University of Notre Dame. He received his B. Tech degree from the Indian Institute of Technology, Delhi and the M.S. and Ph.D. degrees from the California Institute of Technology, all in Electrical Engineering. Prior to joining Notre Dame, he also served as a research associate in the Institute for Systems Research at the University of Maryland, College Park. He received the NSF CAREER award in 2009, and the Donald P. Eckman Award from the American Automatic Control Council in 2013. His research interests include cyber-physical systems, distributed estimation, detection and control, and, in general, the interaction of communication, computation and control.
\end{IEEEbiography}

\end{document}